\documentclass[final,1p,times]{elsarticle}

\usepackage{graphicx}
\usepackage{amsmath, amssymb}
\usepackage{csquotes}
\usepackage{subcaption}
\usepackage{siunitx}
\usepackage{fullpage}
\usepackage{multirow} 
\usepackage{pgfplots} 
\usepackage{bm}
\usepackage{booktabs} 
\usepackage{upgreek}
\usepackage{tablefootnote}
\usepackage{lineno}

\usepackage{hyperref}

\usepackage{chemformula}

\newcommand{\R}{\mathbb{R}}
\newcommand{\N}{\mathbb{N}}
\newcommand{\Z}{\mathbb{Z}}

    \newcommand{\intensityMH}{\lambda_{h}}
    \newcommand{\radiusMH}{r_{h}}
    \newcommand{\MH}{\{S_n\}}

    \newcommand{\generator}[1]{g_{#1}}

    \NewDocumentCommand\grain{mg}{
        \ensuremath{
        G_{#1} \IfNoValueTF{#2}{}{^{(#2)}} 
        }
    }
    
    \newcommand{\distmeasure}[1]{d_{\mathcal{#1}}}
    \newcommand{\seed}[1]{s_{#1}}
    \newcommand{\weight}[1]{r_{#1}}
    \NewDocumentCommand\directionalWeight{gg}{
        \ensuremath{a 
            \IfNoValueTF{#1}{}{_{#1}}}
            \IfNoValueTF{#2}{}{^{(#2)}}
    }

    \newcommand{\millerdirections}[1]{{\langle #1 \rangle}}
    \newcommand{\millerdirection}[1]{{[#1]}}
    \newcommand{\millerplanes}[1]{{\{#1\}}}
    \newcommand{\millerplane}[1]{{(#1)}}
    \newcommand{\orientation}[1]{O_{#1}}
    \newcommand{\inclusion}[1]{I_{#1}}

    \newcommand{\Numgrains}{N_V}
    \newcommand{\NumNeigh}{N_{\text{neigh}}}
    \newcommand{\pNeigh}{p_{\text{neigh}}}
    \newcommand{\grainBound}[1]{\partial \grain{#1}}

    \newcommand{\pincl}{p_{\text{incl}}}

    \newcommand{\habitplane}[1]{H_{#1}}
    \newcommand{\rotAx}[1]{h_{#1}}

    \newcommand{\psplit}{p_{\text{split}}}
    \newcommand{\diam}[3]{\mathsf{diam}_{#1}(#2,#3)}
    \newcommand{\thickness}{\delta}
    
    \newcommand{\depth}{d}
    \newcommand{\halfspace}[1]{\mathcal{H}_{#1}}
    
    \newcommand{\dir}{v}

    \newcommand{\origin}{x}

    \newcommand{\twinningAngle}{\theta}
    \newcommand{\twinningAxis}{\Theta}
    \newcommand{\habitPlane}{E}

\newcommand{\eps}{\mbox{\boldmath$\varepsilon$}}
\newcommand{\sig}{\mbox{\boldmath$\sigma$}}

\newcommand{\xx}{\bm{x}}
\newcommand{\uu}{\bm{u}}
\newcommand{\EE}{\bf{E}}
\newcommand{\CC}{\mathbb{C}}

\newcommand{\nn}{\mbox{\boldmath$n$}}

\journal{Computational Materials Science}

\begin{document}
	
	\begin{frontmatter}
		
		\author{Philipp~Rieder\corref{cor1}\fnref{label1}}
		\author{Matthias~Neumann\fnref{label1}}
		\author{Lucas~Monteiro~Fernandes \fnref{label2}}
		\author{Aude~Mulard\fnref{label3}}
		\author{Henry~Proudhon\fnref{label3}}
		\author{François~Willot \fnref{label2}}
		\author{Volker~Schmidt\fnref{label1}}

		\cortext[cor1]{Corresponding author. E-mail address: philipp.rieder@uni-ulm.de}
		
		\title{Stochastic 3D microstructure modeling of twinned polycrystals for investigating the mechanical behavior of $\gamma$-TiAl intermetallics}
		
		\affiliation[label1]{organization={Institute of Stochastics,  Ulm University},
			city={Ulm},
			postcode={89069}, 
			country={Germany}
		}
		
		\affiliation[label2]{organization={Center for Mathematical Morphology (CMM), Mines Paris, PSL University},
			city={Fontainebleau},
			postcode={12200}, 
			country={France}
		}
		
		\affiliation[label3]{organization={Center of Materials (CMAT), Mines Paris, PSL University},
			city={Evry Cedex},
			postcode={91003}, 
			country={France}
		}
		
		\begin{abstract}
			A stochastic 3D microstructure model for polycrystals is introduced which incorporates two types of twin grains, namely neighboring and inclusion twins. They mimic the presence of crystal twins in $\gamma$-TiAl polycrystalline microstructures as observed by 3D imaging techniques. The polycrystal grain morphology is modeled by means of Voronoi and --more generally--  Laguerre tessellations. The crystallographic orientation of each grain is either sampled uniformly on the space of orientations or chosen to be in a twinning relation with another grain. The model is used to quantitatively study relationships between morphology and mechanical properties of polycrystalline materials. 
			For this purpose, full-field Fourier-based computations are performed to investigate the combined effect of grain morphology and twinning on the overall elastic response.
			For  $\gamma$-TiAl polycrystallines, the presence of twins is associated with  a softer response compared to polycrystalline aggregates without twins. 
			However,
			when comparing the influence on the elastic response, a statistically different polycrystalline morphology has a much smaller effect than the presence of twin grains.
			Notably, the bulk modulus is almost insensitive to the grain morphology and exhibits much less sensitivity to the presence of twins compared to the shear modulus. 
			The numerical results are consistent with a two-scale homogenization estimate that utilizes laminate materials to model the interactions of twins.
		\end{abstract}

		
	\end{frontmatter}
	
	\section{Introduction}\label{sec:Introduction}
	At the engineering scale, most polycrystalline metals or alloys consist of a large number of individual crystalline domains, called grains, each typically ranging in size from \SI{1}{\micro\meter} to a fraction of a millimeter. The mechanical behavior of each grain at the local scale,  characterized by crystallographic orientation, is anisotropic, while at a larger scale, the mechanical response depends on the spatial arrangement of the grains in the microstructure, i.e., on the morphology of the grain system. The latter arises during fabrication and processing, which typically involves heat treatments and solidification~\cite{franccois2012mechanical}. 
	Overall, mechanical properties are influenced by various factors, including inelastic deformation due to the motion of crystalline defects, twinning or phase transformation, as well as brittle or ductile fracture.
	Even in the purely-elastic regime, the local stress-strain state in each grain under mechanical stress is a complex result of the load distribution within the microstructure, depending on both, the crystallographic texture and the morphology of the grain system.
	
	Numerical approaches to model the behavior of polycrystalline materials traditionally rely on  statistical descriptors such as orientation- and misorientation-distribution functions,  as seen in the case of nickel-based superalloys~\cite{groeber2008framework}. Misorientation-distribution functions statistically  quantify the  correlation between crystallographic orientations of neighboring grains,  which is 
	particularly crucial for  metals exhibiting crystal twins. Here the symmetries of the crystal lattice determine the relative orientation of adjacent grains, separated by a twin plane, known as habit plane~\cite{klassen2012mechanical}.
	Frequently, grains in polycrystalline materials are  in multiple twinning relationships, forming so-called twin-related domains. More advanced statistical descriptors  such as \enquote{micro-texture functions}, that describe clustering with respect to crystal orientations~\cite{klassen2012mechanical}, are necessary to model twin-related domains.
	
	A large class of applications involves polycrystals with an overall isotropic mechanical response.  
	This occurs in the important special case where grains exhibit isotropic distribution in shape and crystallographic orientation. Models for realistic polycrystalline materials, therefore, primarily rely on random tessellation of space, with the Laguerre  tessellation being a common choice ~\cite{redenbach2009microstructure,duan2014inverting, furat.2021b}. 
	Note that the Laguerre  tessellation is  a generalization of the Voronoi tessellation, taking the granulometry, or size-distribution, of the resulting grains into account. To account for curved grain boundaries, strong shape anisotropy or multiscale grain-size distribution, 
	alternative approaches like the Johnson-Mehl model and its anisotropic extensions ~\cite{gasnier20153d,gasnier2015fourier} can be used. 
	Another alternative, the generalized balanced power diagram~\cite{alpers.2015, jung2023analytical, sedivy.2016}, offers more degrees of freedom compared to the Laguerre  tessellation. However, the computational cost associated with numerical optimization of morphological descriptors, especially in 3D~\cite{quey2022neper}, limits the use of more general models. A discussion regarding the trade-off between model accuracy and model complexity in terms of morphological descriptors has been provided in \cite{sedivy.2018}.
	Furthermore, in mechanics, numerical results suggest a small effect of the grain morphology on the macroscopic behavior in the elastic regime, whereas crystallography can play an important role, e.g., in the context of thermoelasticity~\cite{gasnier2018thermoelastic}.
	
	In linear elasticity, well-established analytical estimates of the self-consistent type~\cite{berryman2005bounds} are sufficient to estimate the effective properties~\cite{ambos2014a}, even in strongly anisotropic cases and in the presence of cracks~\cite{willot2019thermoelastic}.
	These estimates may also facilitate accurate reconstruction of histograms for strain and stress tensors~\cite{willot2020elastostatic}.
	Mean-field estimates have been used to account for twin-related domains, modeled as individual grains and their effect on the elasto-plastic strain-stress response~\cite{clausen2008reorientation,juan2014double}.
	
	Inclusion twins as well as neighboring twins are present in the TiAl intermetallic family, a material of aeronautical interest, characterized by high stiffness, good corrosion resistance, strength at high temperatures and low density. However, its brittleness and low ductility at room temperature hinder its processing and application in modern components.
	Proper modeling of this material at macroscopic scale requires the understanding of deformation mechanisms at the micromechanical scale, such as dislocation slip and mechanical twinning. These phenomena can be simulated numerically using various techniques, such as finite element analysis and fast Fourier transform (FFT) methods. In this context, the effect on the local mechanical fields of the presence and distribution of  twins in the microstructure is not yet fully understood. This motivates the development of stochastic 3D models for the generation of  polycrystalline microstructures presenting grains in twinning relations, to be used as input for spectral mechanical simulations.
	
	In this paper, the homogenized elastic properties of stochastically generated representative elementary volumes (RVEs) are numerically investigated and compared to analytical results. Two types of tessellations, Voronoi and Laguerre tessellations, are implemented, and the presence (or absence) of twins in  microstructures is explored through three configurations: no twinning, inclusion twinning, and neighboring twinning. The rest of this paper is organized as follows:
	Section~\ref{sec:Experimental} explores the microstructure and material, Section~\ref{sec:Model} describes the stochastic 3D model for polycrystals containing twins, Section~\ref{sec:Mechanics} investigates the elastic response of various polycrystalline models and the effect of twins, and Section~\ref{sec:conclusion} concludes.

	\section{Material microstructure and experimental data}\label{sec:Experimental}
\begin{figure}[h]
		\centering
		\includegraphics[width=\textwidth]{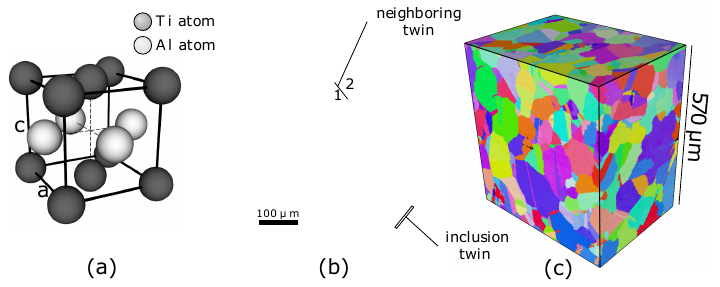}
		\caption{Microstructure of the $\gamma$-TiAl material: (a) L1$_0$ tetragonal crystal structure very close to FCC with a measured ratio $c/a$ of 1.013; (b) EBSD measurement (IPF coloring) of a polished surface of the material showing the numerous twins present (the material has been indexed as cubic here); (c) 3D microstructure reconstruction (IPF coloring) obtained by stacking several hundred EBSD scans after a serial sectioning experiment \cite{TiAl_Tribeam}.}
		\label{fig:gamma_TiAl_EBSD_and_tribeam}
	\end{figure}

	The mechanical properties of binary TiAl alloys can be improved by the addition of heavier elements such as Chromium and Niobium \cite{Kim_JOM_1991} like Ti$_{48}$Al$_{48}$Cr$_2$Nb$_2$, the alloy considered in the present paper. Depending on  heat treatment and  precise composition, this material can adopt very diverse types of microstructures \cite{Dey20091052}. 
	However, the focus of the present paper is on the so-called nearly gamma microstructure, consisting of equiaxed grains of the $\gamma$-phase, with small $\alpha_2$ nodules potentially present but considered insignificant for this analysis.
	The single phase $\gamma$-TiAl is obtained through an isothermal hold just above the eutectoid point ($\sim1130^\circ$C) according to the phase diagram established by McCullough et al. \cite{mccullough_phase_1989} and confirmed by several other authors \cite{kattner_thermodynamic_1992, jones_phase_1993, malinov_experimental_2004}. 
	The obtained $\gamma$-phase exhibits an ordered \emph{Strukturbericht} designation L1$_0$ structure \cite{Kim_JOM_1991},  equivalent to a tetragonal distortion of a face-centred cubic (FCC) structure as shown in Figure \ref{fig:gamma_TiAl_EBSD_and_tribeam}a. The lattice constants are $a=\SI{0.405}{nm}$ and $c=\SI{0.411}{nm}$  as determined by X-ray diffraction measurements \cite{TiAl_Tribeam}, i.e., very close to $a=\SI{0.400}{nm}$ and $c=\SI{0.407}{nm}$ as found in the same material, for instance, in \cite{zambaldi_micromechanical_2010}. The ratio ${c/a=1.013}$ indicates its high similarity to the FCC structure.

	Electron back-scattered diffraction (EBSD) is employed as a technique of choice to characterize the polycrystalline microstructure of metallic materials in detail. The focused ion beam (FIB) interacts with the specimen surface and the resulting scattered electrons form a specific pattern which is captured by a camera and further indexed to measure the local crystallographic  orientation. As a scanning electron microscopy (SEM) technique,  it provides a map of the crystallographic orientations of each point on the scanned surface (represented as  orientation data such as Euler angles or other orientation representations \cite{Rowenhorst2015}). Typically, such data is visualized as 2D cross sections with inverse pole figure (IPF) coloring, as illustrated in Figure \ref{fig:gamma_TiAl_EBSD_and_tribeam}b. In the present study, a detailed reconstruction of the complex $\gamma$-TiAl microstructure is performed, see Figure \ref{fig:gamma_TiAl_EBSD_and_tribeam}c. The data set was obtained by Tribeam tomography performed at the University of California, Santa Barbara~\cite{Echlin_MatChar_2015}, using a FEI Versa microscope equipped with both, a FIB column and a femto-second laser. The TiAl sample was prepared with an approximate cross-section of $500~\upmu\text{m} \times 500~\upmu$m, whereas the laser was employed to ablate 1.5 $\upmu$m of material in the direction perpendicular to the cross-section between each EBSD acquisition. EBSD maps were acquired with  a spatial resolution of 1.5 $\upmu$m  to obtain a volume with an isotropic voxel size  having an edge length of 1.5 $\upmu$m. Further details of the tribeam experiment are reported in \cite{TiAl_Tribeam}. The volume was processed with Dream3D \cite{Dream3d} and several Python routines to obtain the final microstructure with individual segmented grains using an orientation threshold of $5^\circ$.
	
	Stochastic microstructure modeling as described in Section~\ref{sec:Model} is inspired by the observation and processing of the experimental microstructure. 
	Figure \ref{fig:gamma_TiAl_EBSD_and_tribeam}b shows the two considered types of twinned grains, neighboring twinned grains, which are side by side and approximately equal in size, as well as inclusion twinned grains, elongated grains, surrounded by their corresponding \enquote{parent} grains. Note that both types of twins are annealing twins and identical from the crystallographic point of view. However, they are very different in a   morphological sense, which justifies  
	the adopted naming convention. Both types of twins receive specific attention in the  model proposed in the present study, considering morphology and the relationships of crystal orientation between neighboring grains, see Section~\ref{subsec:modelCrystOrisAndTwins}.

	Considering the present $\gamma$-TiAl phase with an L1$_0$ tetragonal crystal structure, closely resembling the FCC structure (often indistinguishable in standard EBSD measurements unless employing high-resolution acquisition with dictionary indexing \cite{Jackson_IMMI_2019}), it is commonly approximated by an FCC lattice. In this data set, the crystal orientation is indexed accordingly.
	This prevents to precisely locate the $c$-axis in each grain but is largely sufficient to determine the type of twin relationship between the grains. 
	
	The 3D data set contains 1966 grains with an average-volume equivalent diameter of 31.37~$\upmu$m. 
	For each grain twinning relations were evaluated, revealing that the vast majority of twins ($\approx 90\%$) are $\Sigma_3$ (see Section \ref{subsec:modelCrystOrisAndTwins} for mathematical  details about twinning relationships) as observed previously in \cite{Gertsman_Scripta_1990}.
	
	\section{Hierarchical stochastic 3D modeling of polycrystalline microstructures with twinning effects}\label{sec:Model}
	
	In this section, we present a stochastic 3D model which has been particularly developed for polycrystalline microstructures with twinning effects as they appear in $\gamma$-TiAl intermetallics described in Section~\ref{sec:Experimental}. The modeling procedure consists of two major steps. First, the initial polycrystalline grain morphology is modeled by random tessellations in the three-dimensional Euclidean space $\R^3$~\cite{chiu.2013}. Second, crystallographic orientations are assigned to the individual grains, including the modeling of neighboring and inclusion twins in the system of grains.
	
	For modeling the initial grain morphology in the first step, we employ Voronoi and --more generally-- Laguerre tessellations. These two types of tessellations are defined by a point pattern $\{s_n\}$ and a marked point pattern $\{(s_n,r_n)\}$, respectively, consisting of so-called generators $g_n=(s_n, r_n)$,  which sometimes will be called seed points of the grains, where $s_n\in\R^3$ and $r_n\in\R$. Thus, we utilize (marked) random point processes to model random tessellations which subdivide the three-dimensional Euclidean space $\R^3$ into non-overlapping sets, each representing a single grain in the microstructure.\footnote{These sets are disjoint, except for their boundaries. This means that the Euclidean space $\R^3$ is subdivided into sets with disjoint interior.} In the case of Laguerre tessellations, the marks associated with the points of the underlying point pattern allow for controlling the size of individual grains. If all marks are equal, the Laguerre tessellation coincides with the special case of a Voronoi tessellation on the same point pattern.

	When modeling crystallographic orientations in the second step, a challenging task is that twin relations between two grains not only depend on their crystallographic orientation, but also on the spatial orientation of their joint grain boundary. More precisely, we have to take into account that the spatial orientation of the grain boundary between a pair of twins coincides with the crystallographic habit plane, see Section \ref{sec:Experimental}.

	\subsection{Modeling grain boundaries by random tessellations}\label{seq:architecture}
	
	We now describe the first step of the modeling procedure, i.e., 
	modeling the polycrystalline grain boundaries. To mimic the structure of polycrystalline materials, random tessellations are used to subdivide the three-dimensional Euclidean space $\R^3$ into sets with disjoint interior, which is common approach in literature~\cite{furat.2021b, sedivy.2016, figliuzzi.2019, seitl.2021}.
	Note that in the context of stochastic geometry, these sets are usually called cells, but in applications to polycrystalline materials it is more convenient to call them grains.
	Each grain $\grain{n}$, $n \geq 1$, consists of all  $x\in\R^3$, which are closer (with respect to a predefined distance function $\distmeasure{T}$) to its generator $\generator{n}$ than to all other generators $\generator{m}$ with $\ m \neq n$.
	Formally,  the $n$-th grain $G_n$ is defined by
	
	\begin{align}
		\grain{n}=\Bigl\{x\in\R^3 \  : \  \distmeasure{T}(x,\generator{n})\leq\distmeasure{T}(x,\generator{m}) \quad \text{ for all } m\neq n\Bigr\}.
		\label{eq:tessellation_general}
	\end{align}
	Motivated by image data representing the polycrystalline structure of $\gamma$-\ch{TiAl} intermetallics, we assume that all grains, except for parent grains of twin inclusions (see Figure \ref{fig:gamma_TiAl_EBSD_and_tribeam}b), are convex and the joint boundaries between two neighboring grains are planar. Due to this assumption, for modeling the initial grain morphology we use random Laguerre and Voronoi tessellations, which share the common feature of producing convex grains with planar boundary segments, i.e. convex polyhedra.
	The randomness of the grain architecture
	is introduced by a randomization of the (marked) point pattern of generators. For this purpose, Mat\'ern hardcore processes are utilized, as described in Section~\ref{sec:point_processes} below. A detailed description of the corresponding random tessellation models is then provided in
	Sections ~\ref{sec:VoronoiTessellation} and ~\ref{sec:isotropic_grain_architecture}.

	\subsubsection{Modeling of seed points by random point processes}\label{sec:point_processes}
	At first we introduce a family of marked point processes which is used in order to model the seed points of grains, i.e.,  the random generators of a tessellation. In particular, for the spatial locations of generators, we use a Mat\'ern hardcore process in $\R^3$, denoted by $\MH$, with intensity $\intensityMH >0$ and hardcore radius $\radiusMH>0,$ see Section~5.4 of~\cite{chiu.2013}. 
	In this model, the generators are a thinning of a homogeneous Poisson point process~\cite{last.2017}, which is performed such that the pairwise distance between generators is larger than the model parameter $\radiusMH>0$. In the case of Laguerre tessellations, each point of the point process $\{S_n\}$ has to be equipped with a random mark. For this, let $R_n$ be a sequence of independent and identically distributed random variables with values in $\R$, which are independent of $\MH$. Then, the  marked point process $\{(S_n,R_n)\}$ induces the random generators of the Laguerre tessellation.
	It is important to note that a Matérn hardcore process is an isotropic point process and leads to  isotropic tessellations. Inducing a Laguerre tessellation by an anisotropic point process instead would lead to an anisotropic tessellation.
	For more details to (marked) point processes, we refer to \cite{chiu.2013,illian2008}. 
	
	\subsubsection{Voronoi tessellation}\label{sec:VoronoiTessellation}
	
	In the case of Voronoi tessellations, the distance function $d_\mathcal{T}$ defining the grain system, see Eq.~\eqref{eq:tessellation_general}, is given by $d_\mathcal{T}:\R^3\times\R^4\rightarrow\R_+$ with $d_\mathcal{T}(x,g)=|x-s|$ for all $x \in \R^3$  and $g=(s,r)\in\R^4$, where $|\cdot|$ denotes the Euclidean distance in $\R^3$.
	Note that the second component $r$ of the generator $g=(s,r)$ 
	is not involved in the definition of  the distance $d_\mathcal{T}$.
	To determine the Voronoi tessellation induced by a given point pattern $s_1, s_2, \ldots\in\R^3$, we plug  $d_\mathcal{T}$ into Eq. \eqref{eq:tessellation_general} and get that
	
	\begin{align*}
		\grain{n}=\Bigl\{x\in\R^3 \  : \  
		|x-s_n|\leq|x-s_m|\quad \text{ for all } m\neq n\Bigr\}.
		\label{eq:tessellation_general}
	\end{align*}
	Less formally, a point $x\in \R^3$ belongs to $G_n$ if the Euclidean distance between $x$ and the seed $s_n$ is less than or equal to the distance between $x$ and all other seeds $s_m$. The grain boundaries consist of points which are equidistant to at least two seeds, see Figure \ref{fig:comparison_vor_lag_gbpd}(a) for the  visualization of a Voronoi tessellation in 2D.

	\subsubsection{Laguerre tessellation}\label{sec:isotropic_grain_architecture}
\begin{figure}[h]
		\centering
		\includegraphics[width=0.74\textwidth]{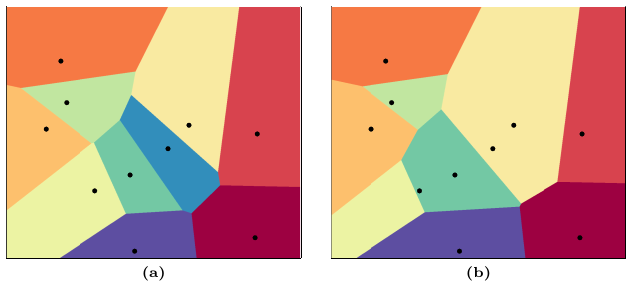}
		\caption{Comparison of different tessellation models in 2D. The black dots indicate the  seed points $\seed{n},n\in\{1,\ldots,10\}$ of the tessellations. (a) Each seed point of a Voronoi tessellation generates a convex grain. Boundaries between two neighboring grains are exactly in the middle between two seeds. (b) A Laguerre tessellation has also convex grains with planar boundary segments. 
			However the joint grain boundary associated with two neighboring seed points is not necessarily between the two seeds,
			see the dark green and lower yellow grains. Moreover, not each generator $\generator{}=(\seed{},\weight{})$ induces a grain (no blue grain). }
		\label{fig:comparison_vor_lag_gbpd}
	\end{figure}

	In comparison to Voronoi tessellations,  Laguerre tessellations allow for more flexibility when modeling polycrystalline materials. In particular, they allow for better controlling the size of the grains through  additive weights. The distance function $d_\mathcal{T}$ leading to the case of Laguerre tessellations is given by ${\distmeasure{T}:\R^3\times\R^4 \rightarrow \R}$ with $\distmeasure{T}(x,g)=|x-\seed{}|^2-r$ for all $x \in \R^3$ and  $g=(\seed{},r) \in \R^4$. Using this distance function for a given marked point pattern $(s_1, r_1), (s_2, r_2), \ldots$, we get that
	
	\begin{align*}
		\grain{n}=\Bigl\{x\in\R^3 \  : \  
		|x-s_n|^2 - r_n \leq|x-s_m|^2 - r_m \quad \text{ for all } m\neq n\Bigr\}.
	\end{align*}
	For generating random marked point patterns, we use a random marked point process $\{(S_n,R_n)\}$, where the points $S_n$ are modeled by a Matèrn hardcore process and the marks $R_n$ are modeled (independently of the points) as independent and identically distributed copies of  $R$,
	a  uniformly distributed random variable on the interval $[r_-,r_+]\subset\R$. A realization of a Laguerre tessellation in 2D is visualized in Figure \ref{fig:comparison_vor_lag_gbpd}(b).

	\subsection{Modeling crystallographic orientations and twinning effects} \label{subsec:modelCrystOrisAndTwins}
	
	In the present paper, given the $\gamma$-TiAl alloy of interest, we restrict ourselves to $\Sigma_3$ twin boundaries which constitute the vast majority of twins experimentally observed  in the material.
	Twin boundaries are specific grain boundaries characterized by a certain misorientation (in the axis/angle sense) between adjacent grains.\footnote{
		In cubic crystal systems, the most prevalent twin boundary is $\Sigma_3$, where one third of the lattice nodes are common between the two grains \cite{Ranganathan_ACtaCrys_1996}.}
	However,  a twinning relation  depends  not only  on the misorientation between two neighboring grains. The spatial orientation of the grain boundary, often referred to as the \textit{habit plane}, is also part of the definition of twins.
	
	The crystallographic nature of the boundary differs depending on the crystal symmetry considered. In the cubic approximation, the rotation is $70.53^\circ$ around one of the $\langle110\rangle$ axes (or equivalently $60^\circ$ around one of the $\langle111\rangle$ axes if the minimal angle representation of misorientations is adopted) \cite{Priester_2013}. In this case the habit plane of the grain boundary is a \{111\} plane with respect to both grains.\footnote{Note that while this is usually the case, one can also find more complex situations where the twin boundary is composed by several parallel segments creating a tortuous boundary that cannot be easily fitted to a plane.} In the tetragonal case (the actual crystal symmetry of the investigated $\gamma$-TiAl), the twin relation is a rotation defined by $\tan(\theta/2) = a/c$\ \footnote{Recall, $a=0.400$nm and $c=0.407$nm denote the lattice constants.} which, in our case, leads to $\theta\approx69.79^\circ$ around either the $\millerdirection{100}$ or $\millerdirection{010}$ direction (which are equivalent with respect to the crystal symmetry). In this case, the habit plane of the grain boundary is either the $\millerplane{011}$ or the $\millerplane{101}$ plane of the tetragonal lattice (which are also equivalent) as reported in Table~\ref{tab:cubicVStetra}.
	
	In Section \ref{sec:Experimental}, two different types of twins were described, namely \textit{neighboring twins} and \textit{inclusion twins}. Due to their differences in morphology, these two types of twins require different modeling techniques. This leads to the following two-step  procedure of twin modeling.
	First, neighboring twins are generated. 
	To introduce a pair of neighboring twins, two adjacent grains are chosen at random. 
	Subsequently, their joint boundary plane is extracted and the orientations of the two grains are set with respect to the conditions of a twinning relation.
	After generating all neighboring twins, the remaining grains are assigned with a random crystal orientation. Afterwards, inclusion twins are generated, where the crystallographic and spatial orientation of an inclusion twin depends on the crystallographic orientation of its parent grain, i.e., the grain in which the inclusion is inserted. 
	
	The present section is divided into several parts. First, in Section~\ref{subsec:twinningrel}, we restrict the stochastic 3D model to a cubic sampling window $V\subset\R^3$. Subsequently, in Section~\ref{seq:isotropy}, we discuss some issues of isotropy. Finally, in Section~\ref{subsec:Algorithmic_realization}, we transfer the introduced model onto a discrete grid $\widehat{V}\subset \Z^3$, which mimics the situation of (discrete) experimental EBSD data, described in Section~\ref{sec:Experimental}. It is important to note that both, the continuous model and its discrete approximation  are defined with periodic boundary conditions. This is consistent with the periodic boundary conditions applied for the mechanical simulations in Section \ref{sec:Mechanics}.
	Furthermore, for referring to crystallographic planes and directions, we utilize common Miller indexing for the continuous model considered in Section \ref{subsec:twinningrel}, whereas for its algorithmic realization in Section \ref{subsec:Algorithmic_realization} we use a more practical notation related to coordinates in Euclidean space.
	
	\begin{table}[h]
		\centering
		\begin{tabular}{c c c c}
			\toprule
			property & symbol  &  cubic symmetry  & tetragonal symmetry\\
			\midrule
			twinning angle & $\twinningAngle$ & $\pi/3 \ (=60^\circ)$ & $2\cdot$tan$^{-1}(c/a) \ (\approx 69.79^\circ)$ \\
			twinning axis & $\twinningAxis$ & $\millerdirections{111}$& $\millerdirection{100} \ , \ \millerdirection{010}$\\
			habit plane & $\habitPlane$ & $\millerplanes{111}$ & $\millerplane{011} \ , \ \millerplane{101}$\\
			\bottomrule
		\end{tabular}
		\caption{Details of the twining relationships depending on the crystal symmetry considered. }
		\label{tab:cubicVStetra}
	\end{table}

	\subsubsection{Model description}\label{subsec:twinningrel}
	
	We now describe the crystallographic aspects of the model, more precisely assigning each grain with
	a crystallographic orientation on a cubic sampling window $V\subset\R^3$. This procedure also incorporates twin relations between adjacent grains into the virtual grain morphology. 
	
	\paragraph{Neighboring twins}\phantom{l} \label{subsec:neighboring twins}
	At first, pairs of adjacent grains to be neighboring twins are determined. For this purpose, we realize a Poisson distributed random variable $\NumNeigh$ with parameter $\pNeigh\cdot\Numgrains/2$,
	conditioned on $\NumNeigh\leq\Numgrains/2$, where $N_V$ denotes the number of grains within the cube $V$ and $p_\text{neigh}\in[0,1]$ is  a model parameter which controls the number of neighboring twins within the virtual microstructure. We then successively choose random pairs of neighboring grains and equip them with a crystallographic orientation,  such that a given  grain cannot be chosen twice. This implies that just twin pairs, and not so-called twin-related domains, which are \enquote{chains} of neighboring twins, are generated in the model.
	
	Thus, the following procedure is repeated until we have either $\NumNeigh$ twin pairs, or no neighboring grains are left, which are not equipped with a crystallographic orientation.
	At first, we choose a grain $\grain{k}$ and one of its neighbors $\grain{\ell}$ at random. 
	Furthermore, let $\grainBound{k,\ell}=\grainBound{\ell,k}=\grain{k}\cap\grain{\ell}$ denote the joint grain boundary of $\grain{k}$ and $\grain{\ell}$. 
	Note that the grain boundaries of all grains, derived by the before mentioned tessellations, consist of planar segments, which implies that $\grainBound{k,\ell}$ is also a planar set (with probability 1).
	By $v_\perp\in\R^3$ we denote a normal vector of $\grainBound{k,\ell}$.\footnote{Note that the following procedure does not depend on the explicit choice of $v_\perp$.}
	
	Recall from Section \ref{sec:Experimental} that for  twins, their joint grain boundary is aligned with a habit plane $\habitPlane$ with respect to the crystal coordinate system of both grains, $\grain{k}$ and $\grain{\ell}$ (for instance in the cubic system, the twin boundary is a $\millerplanes{111}$ plane).
	This property is fulfilled if $v_\perp$ is aligned with the normal of the habit  plane related to $\grain{k}$ and $\grain{\ell}$.
	To include this into the model, a random orientation $\orientation{k}$ is chosen for $\grain{k}$, such that the corresponding habit plane normal is aligned with $v_\perp$.
	The orientation of its twin $\orientation{\ell}$ is unambiguously determined (except for symmetrically equivalent orientations) by  rotating $\orientation{k}$ by 
	the twinning angle $\twinningAngle$ around a twinning axis $ h \in \twinningAxis$, see Table~\ref{tab:cubicVStetra}.
	
\begin{figure}[h]
		\begin{center}
			\includegraphics[width=0.6\textwidth]{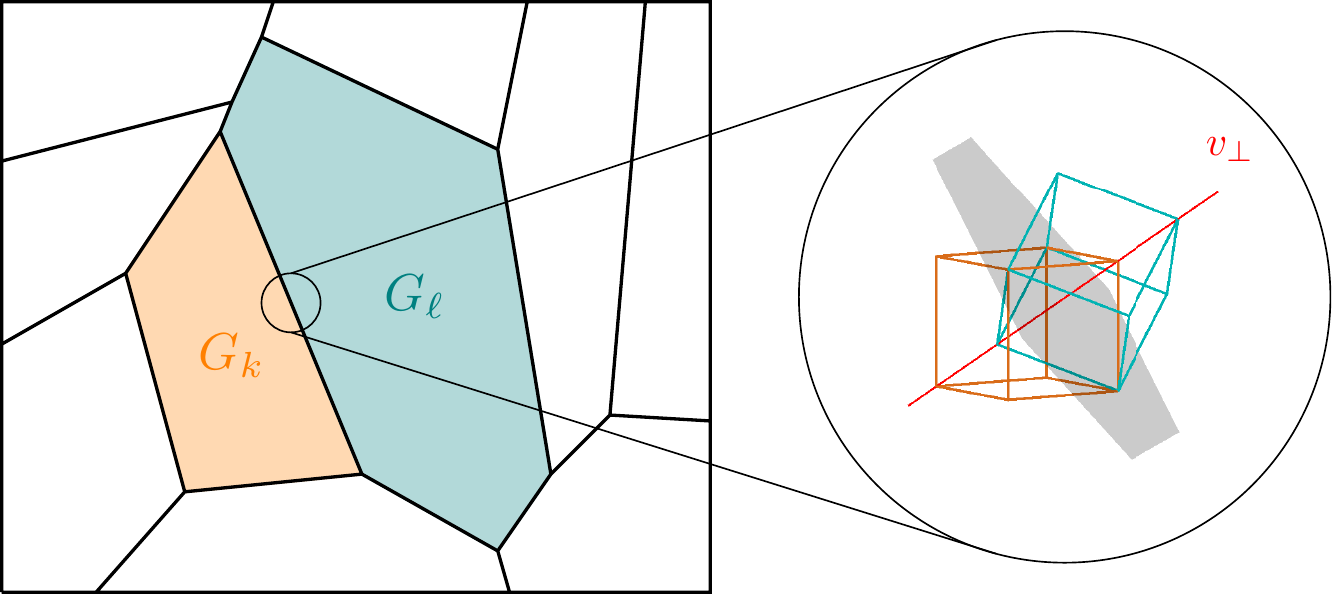}
			\caption{2D sketch of a neighboring twin relation for cubic crystal symmetries. The crystallographic orientations of $G_k$ and  $G_\ell$ are rotated by $\pi/3(=60^\circ)$ around an axis $v_\perp$ (red) to each other. In a morphological sense, $v_\perp$ is normal to the joint grain boundary (gray) of the twinned grains $G_k$ and  $G_\ell$. }
			\label{fig:NeighTwins}
		\end{center}
	\end{figure}

	A 2D visualization of neighboring twins in the case of a cubic symmetry can be found in Figure \ref{fig:NeighTwins}.
	Note that for a given joint grain boundary $\grainBound{k,\ell}$ with its normal $v_\perp$, the orientations of the two adjacent and twinned grains $\grain{k}$ and $\grain{\ell}$ are uniquely defined up to an arbitrary rotation around $v_\perp$ (and symmetrically equivalent orientations).

	\newpage

	\paragraph{Inclusion twins}
	In general, after creating neighboring twins, some grains may not yet be equipped with a crystallographic orientation.
	They are assigned with orientations, which are sampled from the uniform distribution on the space of rotations SO$_3$. Furthermore,
	independently of possible neighboring twin relations, each grain $\grain{}$ can have an inclusion twin $\inclusion{}$ with some  probability $\pincl\in[0,1]$. If grain $G$ is selected to have an inclusion, the following three-step procedure is performed. Otherwise, if no inclusion is generated, the procedure is skipped and  continues with the next grain.
	
	First, the crystallographic orientation $\orientation{\inclusion{}}$ of $\inclusion{}$ is determined by rotating the orientation $\orientation{}$ of the parent grain $G$ around a random rotation axis $\rotAx{} \in \twinningAxis$ at an angle of $\twinningAngle$.
	Furthermore, let $\habitplane{}\in\habitPlane$ denote the corresponding habit plane, which is normal to $\rotAx{}$ for cubic crystal symmetries  and aligned with $\rotAx{}$ in the tetragonal case. 
	Recall that $\habitplane{}$ is not only a crystallographic quantity, but (in a morphological sense) parallel to the joint grain boundary between $\grain{}$ and $\inclusion{}$. To realize this morphological behavior, $\inclusion{}$ is modeled by a dilation of the habit plane.
	
\begin{figure}[h]
		\centering
		\includegraphics[width=\textwidth]{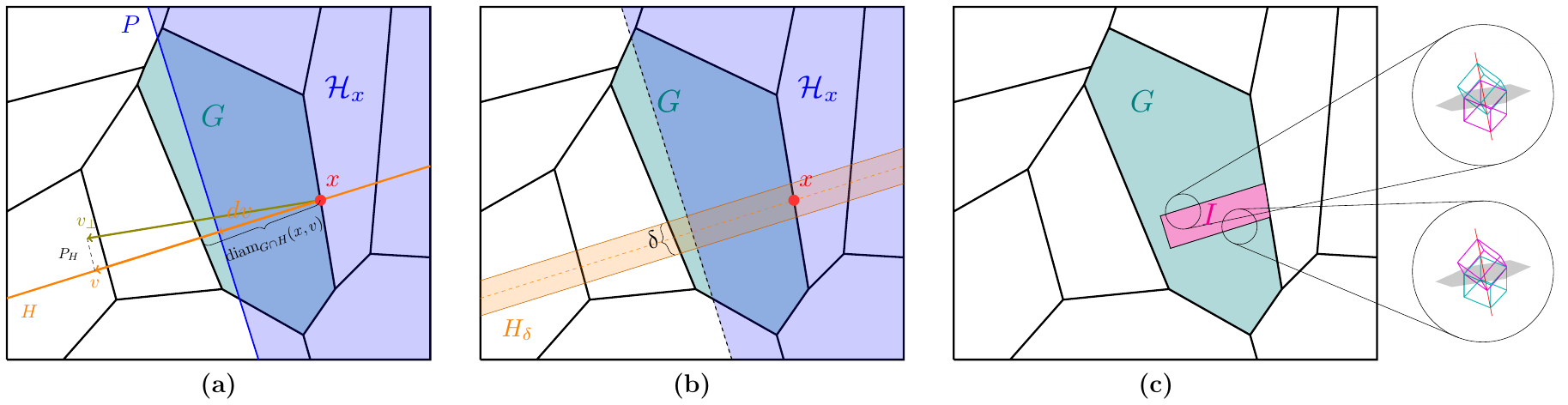}
		\caption{2D sketch of the construction of an inclusion twin relation. (a) 
			Let $\R^3$ be decomposed into two half spaces by the plane $P$, which is normal to $v$ and contains $x+dv$. $\mathcal{H}_\origin$ is defined as the half space containing $x$.
			(b) $H_\delta$ is derived by dilating the habit plane $H$ to a thickness  $\delta$. Intersecting $G$, $H_\delta$ and $\mathcal{H}_\origin$ defines the inclusion $I$. 
			(c) The parent grain $G$ and its inclusion $I$ are in a twin relation at their two main joint grain boundary segments.}
		\label{fig:ConstructInclusion}
	\end{figure}

	The second step, deriving the location and spatial orientation of $\inclusion{}$, is visualized in Figure \ref{fig:ConstructInclusion}(a). To begin with, we choose a random point $\origin $ on $\partial \grain{}$, the grain boundary of $\grain{}$. Without loss of generality we assume that $\origin=o$.  
	Recall that due to the chosen type of  tessellations (Voronoi or Laguerre), the boundary of each grain consists of planar boundary segments. Let $X\subset\R^3$ denote the boundary segment  which contains $\origin$ \footnote{With probability 1, the point $\origin$ is located in the interior of $X$ and not on an edge or vertex between multiple boundary segments.}, and  $v_\perp$ the unit normal vector of $X$, pointing from $\origin$ to the interior of $G$. 
	The \textit{direction of inclusion} $v$ is derived by an orthogonal projection of $v_\perp$ onto $H$, formally expressed by
	\begin{align*}
		v= \frac{P_{\habitplane{}}(v_\perp)}{|P_{\habitplane{}}(v_\perp)|},
	\end{align*}
	where $P_{\habitplane{}}:\R^3\rightarrow \R^3$ denotes the orthogonal projection onto $\habitplane{}$. 
	Note that $v$ is normalized to a length of 1.
	
	In the third step,  
	the shape of $\inclusion{}$ is specified. For this,
	let $\diam{G  }{\origin}{v}$ denote the directional diameter of $G$ at $\origin$ along $v$, which is defined by
	\begin{align*}
		\diam{G  }{\origin}{v} =\sup_{\substack{a\: \in \: c v \cap G\\c \:\in\: \R}}\ \bigl| a-\origin\bigr|.
	\end{align*}
	With some probability $\psplit\in[0,1]$, the grain $G$ is split by $I$ into two separate grains. In the splitting case, the relative depth $D$ of the inclusion equals one. Otherwise, in the case when $G$ is not split, $D$ is sampled from the uniform distribution on the interval $[d_{-},d_{+}]$, where $0< d_{-}< d_{+}\leq 1$. Subsequently, we define the absolute depth as $d=D\cdot \text{diam}_G(\origin,v)$.
	Furthermore, let $\Delta$ denote the relative thickness, sampled uniformly  on $[\delta_{-},\delta_+]$, where $0\leq \delta_{-}< \delta_{+}\leq 1$.  Then, $\delta=\Delta \cdot \rho(G)$ is the absolute thickness of $I$, depending on the volume equivalent diameter $\rho(G)=\sqrt[3]{6/\pi\:\text{vol}_3 (\grain{})}$, where $\text{vol}_3(\, \cdot \, )$ denotes the 3-dimensional Lebesgue measure.

	Finally, $I$ is constructed as the intersection of $G$ with the sets $\mathcal{H}_\origin$ and $ H_\delta$, which are defined as follows.
	Let $P^\prime$ be a plane, normal to $v$ and containing $\origin$. Furthermore, let $P$ be a plane which is parallel to $P^\prime$ and contains $dv$.\footnote{
		Note that $P$ actually contains the point $dv+x$, however due to the choice $x=o$,  $P$ has to contain $dv$.}
	Then, $\mathcal{H}_x$ denotes the half-space resulting from the partition of $\R^3$ by $P$, containing $x$.  
	Formally $\mathcal{H}_\origin$ is defined by
	
	\begin{align*}
		\halfspace{\origin}=\Bigl\{ y\in\R^3:\langle \dir{},y+\depth\dir{} \rangle \geq 0 \Bigr\}.
	\end{align*}
	$ H_\delta$ is the dilated habit plane with a thickness   of $\delta$, 
	mathematically expressed by
	\begin{align*}
		H_\delta=\Biggl\{y\in\R^3: \inf_{z\:\in\: H} |z-y| \leq\frac{\thickness}{2}\Biggl\}.
	\end{align*}
	The construction of $I$ utilizing the half space $\mathcal{H}_\origin$ and the dilated habit plane $ H_\delta$ is visualized in Figure~\ref{fig:ConstructInclusion}(b).
	
	Note that after inserting an inclusion $I$  into a grain $G$, the  grain architecture of the initial tessellation  is changed, i.e., a new grain is introduced (the inclusion $I$). In particular, a grain $G$ might (if the inclusion twin spans over the whole grain $G$) also be split into two new grains (without changing the crystallographic orientation), which corresponds well to the experimentally observed microstructures.
	
	\subsubsection{Isotropy}\label{seq:isotropy}

	Since an isotropic microstructure is  assumed  for the mechanical simulations in Section \ref{sseq:linearElasticProblem}, it is noteworthy that neither the generation of neighboring twins nor inclusion twins affects the isotropy of microstructure in terms of crystallography and morphology. Recall that the crystallographic orientation of neighboring twins depends on the spatial orientation of their joint grain boundary. Since Voronoi and Laguerre tessellations are isotropic tessellation models\footnote{At least, if the underlying point process is isotropic like in the present paper.}, the spatial orientation of the grain boundaries is also isotropic. This implies that the choice of neighboring twins is invariant under rotations and thus isotropic. Grains not assigned with a crystallographic orientation during the neighboring twin procedure, are equipped with a  crystallographic orientation sampled from the uniform distribution on the space of rotations SO$_3$, which also  does not violate the isotropy. Additionally, the crystallographic orientation of inclusion twins is determined  on the basis of crystallographic orientation of the parent grain. Consequently, the overall isotropy is not affected either.

	\subsubsection{Implementation and discretization of the stochastic 3D model} \label{subsec:Algorithmic_realization}
	The stochastic 3D model for polycrystalline microstructures introduced in Section \ref{subsec:twinningrel} is defined in the three-dimensional Euclidean space $\R^3$. As geometry input for the numerical computation of mechanical properties in Section~\ref{sec:Mechanics}, discrete model realizations are generated on a cube $V\subset \R^3$ intersected with the voxel grid, i.e. on $V \cap \Z^3$. Thus, in this section, an algorithmic description for the generation of discrete model realizations is provided. Recall that the image data obtained by EBSD measurements is also given as voxelized data, see Section~\ref{sec:Experimental}.
	
	For each quantity considered in Section \ref{subsec:twinningrel} its discrete  counterpart is marked by a hat. For instance, a grain $G$ is approximated by its voxel representation $\widehat{G}=G\cap\widehat{V}$, where $\widehat{V}=V\cap {\Z}^3$.
	Furthermore, we use the notation $\mathcal{N}_m(z)$ for the $m$- neighborhood of a voxel $z\in\widehat{V}$ with $m=18$ or $m=125$.\footnote{Formally, these neighborhoods are defined as $\mathcal{N}_{18}(z) = \bigl\{x \in \Z^3: 0 < |z-x| \leq \sqrt{2}\, \bigr\}$ and $\mathcal{N}_{125}(z) = \bigl\{x \in \Z^3: |z-x| < 3 \bigl\}$.}
	Note that we use periodic boundary conditions and thus each voxel has the same number of neighbors. We also make use of different representations of rotations, namely matrix representations and axis-angle representations. For more details, we refer to \cite{Rowenhorst2015}. 
	Furthermore, we use the convention that a crystallographic orientation $O$ consists of a crystal symmetry (in the present paper either cubic or tetragonal)  as well as a rotation $R\in$SO$_3$.

	\paragraph{Neighboring twins}
	Let $\widehat{G}_k$ and $\widehat{G}_\ell$ denote the voxel representations of a pair of arbitrarily chosen neighboring grains $G_k$ and $G_{\ell}$, which are assumed to share a $\Sigma_3$ twin boundary. Recall that the crystallographic orientations of $\widehat{G}_k$ and $\widehat{G}_\ell$ depend on the spatial orientation of their joint grain boundary $\partial\widehat{G}_{k,\ell}$ as it aligns with its crystallographic habit plane.
	For Laguerre tessellations in the Euclidean space $\R^3$, the joint grain boundary $\partial G_{k,\ell}$ between two neighboring grains $G_k$ and $G_\ell$ is normal to the vector connecting their seed points $s_k$ and $s_\ell$. Thus,
	by defining $v_\perp = s_k - s_\ell$ we obtain the  normal $v_\perp$ of $\partial G_{k,\ell}$, which is also used as the normal $\hat v_\perp$ corresponding to the discrete grain boundary $\partial\widehat{G}_{k,\ell}$.
	
	In accordance with the description of neighboring twins  $G_k$ and $G_{\ell}$ given in Section~\ref{subsec:twinningrel},
	the crystallographic orientation $R_k$ of $G_k$ is determined by  the composition $R_k=R^\prime\circ\Tilde{R}\circ R_v$. Here,  $R_v=[v_\perp,v_1,v_2]$ is an orthonormal basis (ONB) of $\R^3$, chosen such that $v_1$ and $v_2$ are spanning  $\partial G_{k,\ell}$. 
	Because the vectors $v_1$ and $v_2$ are deterministically computed\footnote{Consider the vector $v_1 = (-v_{\perp,2}, v_{\perp,1}, 0)^\intercal$, where $v_{\perp,i}$ represents the $i$-th component of $v_\perp$. The vector $v_2$ is determined as the cross product between $v_\perp$ and $v_1$. To establish an ONB, it is essential to normalize these vectors to a length of 1.} based on $v_\perp$, the rotation matrix $R_v$ has to be randomly rotated around $v_\perp$ such that $R_k$ does not depend on the particular choice of the ONB defining $R_v$. This is accomplished by utilizing the 
	rotation $\Tilde{R}$, the axis-angle representation of which is given by $(v_\perp,\Tilde{\omega})$, where $\Tilde{\omega}$ is sampled from the uniform distribution on $[0,2\pi)$. 
	Subsequently, the composition $\Tilde{R}\circ R_v$ is aligned such that $v_\perp$ coincides with the normal of the habit plane concerning the crystal symmetry by applying $R^\prime$.
	By composing $R_\parallel$ with $R_k$, we derive $R_\ell$, the underlying rotation of $O_\ell$.
	Finally  $R_k$ and $R_\ell$ 
	are transformed by a random rotation that is equivalent with respect to the crystal symmetry, by applying $R_\text{equi}(R_j), \ j=k,\ell$. We
	refer to Table \ref{tab:cubicVstetraNBtwins} for the specific construction rules for each of these rotations.

	\begin{table}[h]
		\centering
		\begin{tabular}{c|  c |c}
			\toprule
			&face centered cubic symmetry  & face centered tetragonal symmetry\\
			\midrule
			grain boundary $\partial G_{k,\ell}$  &\multicolumn{2}{c}{$R_v=[v_\perp,v_1,v_2]$,}\\
			and its normal&\multicolumn{2}{c}{
				$R_v$ ONB and $v_\perp $ normal to $\partial G_{k,\ell}$}\\
			\midrule
			random rotation around $v_\perp$&\multicolumn{2}{c}{$\Tilde{R}=\bigl(v_\perp,\Tilde{\omega}\bigr)$, \ \ with $\Tilde{\omega}\sim\mathcal{U}\bigl([0,2\pi)]\bigr)$}\\
			\midrule
			\multirow{3}{*}{alignment with habit plane } &
			\multirow{3}{*}{$R^\prime=\biggl(v_1-v_2,-\operatorname{cos}^{-1}\Bigl(\sqrt{2/3}\,\Bigr)+\pi/3      \biggr)$} & 
			$R^\prime=\bigl(\Tilde{R}\circ R_v \circ (0,1,0)^\intercal, \omega^\prime\bigr)$\\
			& & $\omega^\prime =(-1)^i \operatorname{tan}^{-1}(a/c), \ i \sim \mathcal{U}\bigl(\{ 0,1 \}\bigr)$\\
			& &$a,c$ lattice constants\\
			\midrule
			rotation matrix of $O_k$&\multicolumn{2}{c}{$R_k=R^\prime\circ\Tilde{R}\circ R_v$}\\
			\midrule
			\multirow{2}{*}{twin relation}&\multirow{2}{*}{$R_\parallel=\Bigl(R_k\circ\sqrt{1/3}\,\bigl(1,1,1\bigr)^\intercal,\pi/3\Bigr)$} &  $R_\parallel=\bigl(R_k\circ(0,1,0)^\intercal,\omega_\parallel\bigl)$\\
			&&$\omega_\parallel=(-1)^{i+1} 2\operatorname{tan}^{-1}(a/c) $ with $i$ of $R^\prime$\\ 
			\midrule
			rotation matrix of $O_\ell$&\multicolumn{2}{c}{$R_\ell=R_\parallel \circ R_k$}\\
			\midrule
			identification with equivalent 
			&\multicolumn{2}{c}{\multirow{2}{*}{$R_{\text{equi}}(R)=R_{100}(R)\circ R_{001}(R)$}}\\
			rotation\\
			\midrule
			rotation around $\millerdirection{100}$&\multicolumn{2}{c}{$R_{100}(R)=\bigl(R\circ(0,0,1)^\intercal,i\pi/2\bigr)$ \ \ $i\sim\mathcal{U}\bigl(\{0,1,2,3\}\bigr)$}\\
			\midrule
			rotation around $\millerdirection{100}$&\multicolumn{2}{c}{$R_{001}(R)=\bigl(R\circ(1,0,0)^\intercal,i\pi\bigr)$ \ \ $i\sim\mathcal{U}\bigl(\{0,1\}\bigr)$}\\
			\bottomrule
		\end{tabular}
		\caption{Notation used to construct the orientations of neighboring twins. The rotations are presented in the form of matrices, compositions of matrices, and as axis-angle tuples. Note that 
			$( \cdot\, , \cdot\, , \cdot)$ indicates a vector in $\R^3$ and not a crystal plane in Miller indices. For the sake of readability, we denote a uniformly distributed random variable $X$ on a set $S$ as $X\sim\mathcal{U}(S)$. }
		\label{tab:cubicVstetraNBtwins}
	\end{table}

	\paragraph{Inclusion twins}
	Before generating discretized inclusion twins, recall that each grain $G$ not assigned with a crystallographic orientation in the previous step, is equipped with an orientation sampled from the uniform distribution on SO$_3$.
	For this we use the statistical software package R (version 4.2.2) and the package rotations \cite{Stanfill2014} to sample from the uniform distribution on SO$_3$.
	
	The precise definitions of quantities related to the crystallographic orientation are not explicitly given in the following paragraph. Instead, these definitions are provided  in Table~\ref{tab:cubicVstetraNBtwins} for cubic and tetragonal crystal symmetries.
	
	To model a discretized inclusion $\widehat{I}$ of $  \widehat{G}$, a random twinning axis $  \widehat{\Theta^\prime}$ is simulated relative  to the crystal symmetry of $  \widehat{G}$. The orientation $O_I$ of $  \widehat{I}$ is then determined by the rotation $R_I=(  \widehat{\Theta},\theta)$, 
	where $\widehat{\Theta}=R_G\circ\widehat{\Theta^\prime}$ with $R_G$, the rotation corresponding to the orientation $O_G$ of $\widehat{G}$.
	Similar to generating  neighboring twins, the orientation of $I$ has to be identified with an equivalent rotation by applying $R_\text{equi}(R_I)$, see Table \ref{tab:cubicVstetraNBtwins}. 
	Note that the spatial orientation of the discrete habit plane $\widehat{H}$ is directly determined from $R_I$ and $R_G$ (Table \ref{tab:cubicVstetraNBtwins}). Furthermore, $\widehat{H}$ is represented by its normal $\hat{h}$, where 
	$R_I$ and $\hat{h}$ are analytically exact and equal to the corresponding quantities in the continuous model.

	After establishing the crystallographic orientation of the inclusion $\widehat{I}$, the subsequent steps concern its spatial arrangement. Recall that in
	contrast to  neighboring twins, the construction of inclusions requires a modification of the morphology of the grain, which was initially generated by a Voronoi or Laguerre tessellation.
	Unlike the continuous representation of inclusion twins described in Section \ref{subsec:twinningrel}, now we are  limited to a voxelized representation of the grain architecture, which requires a discrete approximation of the inclusion on the voxel grid.
	
	Let $\partial\widehat{G}=\{ x\in \widehat{G} : \text{ there exists a } y\in\mathcal{N}_{18}(x) \text{ with } y\notin \widehat{G} \}$ be the boundary of $\widehat{G}$, i.e. $\partial\widehat{G}=\{x_1,\ldots,x_N\}\subset\widehat{V}$  for some integer $N\ge 1$.  Furthermore, let $\widehat{X}=x_j$  be a random voxel on $\partial\widehat{G}$, where $j$ is sampled from the uniform distribution on $\{1,\ldots,N\}$.
	To get the normal $  \hat{v}_\perp$ of $\partial   \widehat{G}$ in $\widehat{X}$, we perform a principal component analysis on the point cloud $\bigl\{ x\in\mathcal{N}_{125}\bigl(\widehat{X}\bigl) : x \in \partial\widehat{G} \bigr\} $. Doing so, we obtain the three components $v_1,v_2$ and $  \hat{v}_\perp$
	, see e.g. ~\cite{Hastie2009}.
	The vectors $v_1,v_2$ indicate the two main directions of the extension of the point cloud, i.e. the local grain boundary $\partial \widehat{G}$ in $\widehat{X}$. The vector  $\hat{v}_\perp$ is normal to $v_1$ and $v_2$. Without loss of generality, we assume that $  \hat{v}_\perp$  is pointing into $\widehat{G}$.
	The direction of the inclusion $  \hat{v}$ is derived by projecting $  \hat{v}_\perp$ orthogonal onto the habit plane $\widehat{H}$ by 
	
	\begin{align*}
		\hat{v}=  \hat{v}_\perp-\frac{\bigl\langle   \hat{h},  \hat{v}_\perp\bigr\rangle}{\bigl|\,  \hat{h}\,\bigl|} \,  \hat{h},
	\end{align*}
	where $  \hat{h}$ denotes the normal of $\widehat{H}$. Without loss of generality, we assume that $  \hat{v}$ is a unit vector of length $1$. 
	The approximated diameter of $\widehat{G}$ in $\widehat{X}$ along $  \hat{v}$ is defined by 
	
	\begin{align*}
		\widehat{\text{diam}}_{\widehat{G}}\,\bigl(\widehat{X},  \hat{v}\bigr)=\max\Bigl\{i\in\N : \lceil \, i   \hat{v} \, \rfloor + \widehat{X} \in \widehat{G}\, \Bigr\},
	\end{align*}
	where $\lceil \, \cdot \, \rfloor$ denotes rounding to the closest integer. 
	Furthermore, let  $  \hat{\rho}(\widehat{G})=\sqrt[3]{6 \, \#\widehat{G} / \pi }$ be the volume equivalent diameter, where $\#\widehat{G}$ denotes the cardinality of $\widehat{G}$.
	The absolute depth $  \hat{d}$ and thickness $  \hat{\delta}$ are determined analogously to their continuous counterparts described in Section~\ref{subsec:neighboring twins}. Finally, we are able to define  $\widehat{I}=\widehat{G}\cap\widehat{\mathcal{H}}_{\widehat{X}}\cap\widehat{H}_\delta$, where
	
	\begin{align*}
		\widehat{\mathcal{H}}_{\widehat{X}}&=\Bigl\{ x \in \widehat{V} : \bigl\langle   \hat{v},x+\hat{d} \, \hat{v}\bigr\rangle \geq 0 \Bigr\} \quad
		\text{ and } \quad
		\widehat{H}_{\hat{\delta}}=\Biggl\{ x\in \widehat{V} : \inf_{y\in   \widehat{H}} \bigl| x-y \bigr|\leq \frac{  \hat{\delta}}{2} \Biggr\}.
	\end{align*}
	
	\begin{table}[h]
		\centering
		\begin{tabular}{c|  c |c}
			\toprule 
			&face centered cubic symmetry  & face centered tetragonal symmetry\\
			\midrule
			vector representation &$  \widehat{\Theta}^\prime\in\Bigl\{\frac{1}{\sqrt{3}}\bigl(1,1,1\bigr)^\intercal,\frac{1}{\sqrt{3}}\bigl(-1,1,1\bigr)^\intercal,$ & 
			\multirow{2}{*}{$  \widehat{\Theta}^\prime \in \bigl\{(1,0,0)^\intercal,(0,1,0)^\intercal\bigr\}$}\\
			of twinning axis&\phantom{asdsdsdf}$\frac{1}{\sqrt{3}}\bigl(1,-1,1\bigr)^\intercal,\frac{1}{\sqrt{3}}\bigl(1,1,-1\bigr)^\intercal\Bigr\}$ & \\
			\midrule   
			twinning axis in &\multicolumn{2}{c}{\multirow{2}{*}{$  \widehat{\Theta}=R_G\circ  \widehat{\Theta}^\prime$}}\\
			reference system\\
			\midrule 
			\multirow{3}{*}{twinning angle}&\multirow{3}{*}{$\theta = \pm\frac{\pi}{3}$} & $\theta=(-1)^i \ 2 \operatorname{tan}^{-1}\Bigl(\frac{a}{c}\Bigr)$\\
			&& $i\sim\mathcal{U}\bigl(\{0,1\}\bigr)$\\
			&& a,c lattice constants\\
			\midrule
			rotation matrix of $O_I$&\multicolumn{2}{c}{$R_I=\bigl(  \widehat{\Theta},\theta \bigr)$}\\
			\midrule
			$\begin{array}{l}
				\text{habit plane normal}\\
				\text{in reference system}
			\end{array}$&&
			$  \hat{h}^\prime=\begin{cases}
				\Bigl(0,\frac{a}{c},\text{sgn}(\theta)\Bigr)^\intercal \ \ \text{ if } \   \widehat{\Theta}^\prime=(1,0,0)^\intercal,\\
				\Bigl(\frac{a}{c},0,-\text{sgn}(\theta)\Bigr)^\intercal \ \ \text{ else.} 
			\end{cases}$\\
			\midrule
			habit plane normal&$  \hat{h}=  \widehat{\Theta}$&$  \hat{h}=R_G\circ  \hat{h}^\prime$\\
			\bottomrule
		\end{tabular}
		\caption{Notation used to construct the orientations of inclusion twins. The rotations are presented in the form of matrices, compositions of matrices, and as axis-angle tuples. Note, that the for the representation of axes, $( \cdot\, , \cdot\, , \cdot)$ indicates a vector in $\R^3$ and not a crystal plane in Miller indices. For the sake of readability, we denote a uniformly distributed random variable $X$ on a set $S$ as $X\sim\mathcal{U}(S)$.}
		\label{tab:cubicVstetraIncltwins}
	\end{table}

	\subsection{Specification of model parameters}\label{seq:ModelParameters}
	The study of different microstructures  in Section \ref{sec:Mechanics} is carried out on the voxel representation $\widehat{V}$ of a cubic sampling window  $V$ with an edge length of $512$ voxels. Recall that the generated structures adhere to periodic boundary conditions.
	To generate microstructures according to the stochastic 3D model outlined in Sections \ref{seq:architecture} and \ref{subsec:modelCrystOrisAndTwins} we have to initially generate  a random  pattern of seed points by a Matérn hardcore process. This process is defined by a hardcore radius $r_h=5$ and an intensity $\lambda_h = 15360/512^3(\approx 1.14\cdot 10^{-4})$ which corresponds to an expectation of $15360$ seed points 
	within the sampling window $V$.
	While a Voronoi tessellation does not need any  additional information, the additive weights of the Laguerre tessellation are sampled from the uniform distribution on the interval $[-16,16]$.
	
	For both tessellation types, we consider three different configurations of twinning parameters.  While these configurations are typically not observed in experimental data, they serve as extreme cases of twinning occurring in realistic materials. 
	(i) \enquote{Neighboring twins} (NT), where we put $p_{\text{neigh}}=1, p_{\text{incl}}=p_{\text{split}}=0$.  Furthermore,
	the number of neighboring twin pairs $N_{\text{neigh}}$ is drawn from a Poisson distribution with parameter \mbox{$p_{\text{neigh}} \cdot N_V/2= N_V/2$}, where $N_V$ denotes the number of grains in $\widehat{V}$, while ensuring $N_{\text{neigh}}\leq N_V$. Notably, when $p_{\text{neigh}}\approx 1$ (particularly in this twin configuration) there are less than $N_\text{neigh}$ twin pairs established in general. The latter occurs due to consistency reasons as it is rather unlikely that such a high predefined number of neighboring grains can be realized in the algorithm.
	(ii) \enquote{Inclusion twins} (IT), where we put $p_{\text{neigh}}=0, p_{\text{incl}}=p_{\text{split}}=1$, i.e., 
	each grain $G$ within the tessellation is assigned with an inclusion twin, effectively splitting $G$ into two distinct parts. 
	(iii) \enquote{Without twins} (WT), where we put $p_{\text{neigh}}=p_{\text{incl}}=p_{\text{split}}=0$, i.e., 
	the microstructure does not exhibit any twins.
	
	The relative thickness $\Delta$ of an inclusion is sampled from the uniform distribution on $[\delta_-,\delta_+]=[0.05,0.1]$. 
	Note, that the interval $[d_-,d_+]$ is not further specified, because a splitting probability $p_\text{split}=p_\text{incl}\in\{0,1\}$ leads to a relative depth $D\in\{0,1\}$.

	Altogether, a total of 12 scenarios is investigated in Section \ref{sec:Mechanics} consisting of two types of tessellations (Voronoi and Laguerre), three twinning configurations (NT, IT and WT) and two crystal symmetries.

	\section{Effect of twinning and grain morphology on the elastic response}\label{sec:Mechanics}
	In this section, mechanical properties of $\gamma$-TiAl polycrystals with and without twins are investigated in the elastic regime.
	To address this problem, numerical computations are carried out utilizing 
	a spectral solver, based on fast Fourier transforms (FFT), 
	proposed by Moulinec and Suquet~\cite{moulinec_fast_1994}. 
	These methods rely on an implicit integral equation for the strain field
	and the associated Green's operator. FFT schemes are a class of computationally efficient algorithms designed to compute the physical response (mechanical, thermo-mechanical, conductive) of heterogeneous media under macroscopically periodic boundary conditions. Applying this method to linear elastic theory, strain and stress tensors are defined on a regular grid, i.e. on the voxel grid. Coupled to modern experimental techniques such as 3D microtomography and FIB imaging, this approach is powerful in apprehending not only homogenized properties but also local fields~\cite{Escoda15b} 
	and offering substantial  numerical speedup compared to finite element methods~\cite{Koishi2017}.

	For the sake of simplicity, we adopt infinitesimal strain theory in the present study. Moreover, a discrete Green's operator is applied, which is  consistent with centered differences on a rotated grid together with the \enquote{direct scheme} presented in~\cite{willot_fourier-based_2015}. 
	Alongside numerical computations, 
	we present self-consistent type estimates for polycrystals containing crystallographic twins. These estimates are explicitly derived and compared to FFT predictions.
	
	\subsection{Linear elastic problem and FFT-based computations} \label{sseq:linearElasticProblem}
	Consider the following linear elasticity problem:
	
	\begin{equation}\label{eq:elasticproblem}
		\sig(\xx)=\CC(\xx):\eps\bigl(\uu(\xx)\bigr) , \qquad
		\text{div}(\sig)=0, \qquad \xx\in V,
	\end{equation}
	where 
	$\sig(\xx)$ denotes the Cauchy stress tensor,
	$\CC(\xx)$ is the stiffness tensor of the local crystal
	and $\eps(\uu(\xx))=\text{grad}_{\text{sym}}(\uu(\xx))$, denoted as Green-Lagrange strain tensor, equals the symmetric gradient of $\uu(\xx)$.
	The strain and stress fields are defined on a cubic domain $V\subset{\R^3}$,
	and extended, by periodicity, to $\mathbb{R}^3$.
	Equivalently, by imposing periodic boundary conditions on the boundary $\partial V$ of $V$, the strain field $\eps(\uu(\xx))$ can be split into a prescribed average $\EE$ and an unknown periodic term $\eps(\uu^\ast(\xx))$ given by 
	
	\begin{equation*}\label{eq:epssplit}
		\eps\bigl(\uu(\xx)\bigr)=\EE+\eps\bigl(\uu^\ast(\xx)\bigr), \quad
		\bigl\langle \eps(\uu) \bigr\rangle=\EE, \quad \uu^*(\xx)=\uu(\xx)-\EE\cdot\xx,
	\end{equation*}
	where $\uu^*$ denotes the periodic part of the displacement field,
	and  $\langle \, \cdot \,  \rangle$  spatial averaging over $V$. Furthermore, we prescribe
	\begin{equation*}\label{eq:perBCs}
		\uu^\ast\:\#, \quad
		\sig\cdot\bf{n}\:\textrm{--}\,\#,
	\end{equation*}
	where $\#$ denotes a $Q$-periodic field and $\textrm{--}\,\#$ an anti-periodic field. 
	The effective properties are defined by the overall stiffness tensor $\CC^*$ defined by
	
	\begin{equation*}\label{eq:eff}
		\bigl\langle\sig(\xx)\bigr\rangle = \CC^*:\EE.
	\end{equation*}
	Referring to Section \ref{seq:isotropy}, it is important to note that the considered models are statistically isotropic, implying isotropy of the tensor $\CC^*$~\cite{ambos2014a}.
	For periodic media, this condition is satisfied asymptotically when the domain $V$ surpasses the size of the grains significantly and can be considered as representative volume element (RVE)~\cite{lantuejoul1991ergodicity}. This was numerically verified for the presented model via FFT computations.

	A macroscopic strain field is applied along $6$ independent strain directions enabling to compute all $21$ components of the stiffness tensor $\CC^*$.
	Furthermore, we define the effective bulk modulus $\kappa$ and shear modulus $\mu$ as the closest isotropic elastic tensors in the Euclidean sense~\cite{moakher_closest_2006, gazis_elastic_1963, fedorov_theory_1968}.
	
	\begin{table}[h]
		\begin{center}
			\begin{tabular}{ c|c|c|c|c|c|c } 
				\toprule
				symmetry & $c_{11}$ & $c_{12}$ & $c_{13}$ & $c_{33}$ & $c_{44}$ & $c_{66}$ \\
				\midrule
				\multirow{1}{4em}{cubic} & $183$ & $74$ & $-$ & $-$ & $105$ & $-$  \\
				\multirow{1}{4em}{tetragonal} & $183$ & $74$ & $74$ & $178$ & $105$ & $78$ \\
				\bottomrule
			\end{tabular}
			\caption{Stiffness constants in GPa for cubic and tetragonal crystal symmetries.}
			\label{table:elastic_values}
		\end{center}
	\end{table}
	
	Utilizing single-crystal constants from~\cite{fu_elastic_1990,tanaka_single-crystal_1996}, the stiffness tensors of Hooke's law~\eqref{eq:elasticproblem} for cubic $\CC^C$ and tetragonal $\CC^T$ crystal symmetries are given by
	
	\begin{equation*}
		\CC^C=\left(\begin{tabular}{cccccc}
			$c_{11}$ & $c_{12}$     & $c_{12}$ &          &          &           \\
			& $c_{11}$     & $c_{12}$ &          & $0$      &           \\
			&              & $c_{11}$ &          &          &           \\
			&              &          & $2c_{44}$ &          &           \\
			& \textit{sym} &          &          & $2c_{44}$ &           \\
			&              &          &          &          &  $2c_{44}$
		\end{tabular}\right), \qquad
		\CC^T=\left(\begin{tabular}{cccccc}
			$c_{11}$ & $c_{12}$     & $c_{13}$ &          &          &           \\
			& $c_{11}$     & $c_{13}$ &          & $0$      &           \\
			&              & $c_{33}$ &          &          &           \\
			&              &          & $2c_{44}$ &          &           \\
			& \textit{sym} &          &          & $2c_{44}$ &           \\
			&              &          &          &          &  $2c_{66}$
		\end{tabular}\right)
	\end{equation*}
	where the corresponding stiffness constants $c_{ij}$ are presented in Table \ref{table:elastic_values}. Moreover, the values of $\sig$ and $\eps$, expressed in Voigt conventions, are
	
	\begin{equation*}
		\sig=\Bigl(\sigma_{11}, \sigma_{22}, \sigma_{33}, \sqrt{2}\sigma_{23}, \sqrt{2}\sigma_{13}, \sqrt{2}\sigma_{12}\Bigr)^\intercal, \quad
		\eps=\Bigl(\varepsilon_{11}, \varepsilon_{22}, \varepsilon_{33}, \sqrt{2}\varepsilon_{23}, \sqrt{2}\varepsilon_{13}, \sqrt{2}\varepsilon_{12}\Bigr)^\intercal.
	\end{equation*}
	The numerical computations are carried out 
	on three realizations of each model, namely Laguerre and Voronoi  with the three twinning configurations neighboring twins, inclusion twins  and without twins (Section~\ref{seq:ModelParameters}), which results in 18 samples.

	\subsection{Self-consistent estimates}
\begin{figure}[h]
		\centering
		\includegraphics[width=0.7\textwidth]{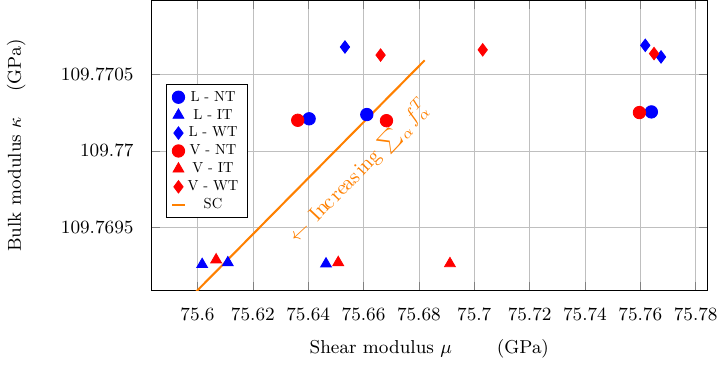}
		\caption{Effective elastic response of polycrystals with tetragonal elastic symmetry class.
			FFT results:  Laguerre (L, blue symbols) and Voronoi models (red symbols), with neighboring twins (circles), inclusion twins (triangles),  and without twins (diamonds). The orange line represents self-consistent estimates for varying twinned volume fraction. }
		\label{fig:tetra_results}
	\end{figure}

	As expected, the Voigt and Reuss bounds,
	also known as Hill bounds
	in the context of polycrystals~\cite{berryman_bounds_2011},
	significantly overestimate or underestimate the elastic moduli.
	Likewise, Hashin-Shtrikman (or Hashin-Shtrikman-Walpole)
	bounds offer only marginal improvements. 
	For instance, considering the stiffness tensor for tetragonal elastic symmetry class,
	the Hill bounds provide $\SI{109.760}{GPa}\leq \kappa \leq \SI{109.778}{GPa}$  and $\SI{72.3}{GPa}\leq \mu\leq \SI{79.1}{GPa}$  for the
	bulk modulus $\kappa$ and shear modulus $\mu$, whereas Hashin-Shtrikman bounds are slightly more narrow, e.g. $\SI{72.4}{GPa}\leq \mu\leq \SI{79.0}{GPa}$.
	The determined bulk and shear moduli are shown in Figure~\ref{fig:tetra_results}.
	
\begin{figure}[h]
		\centering
		\includegraphics[clip, trim=3.5cm 15cm 9cm 6.3cm, width=0.25\textwidth]{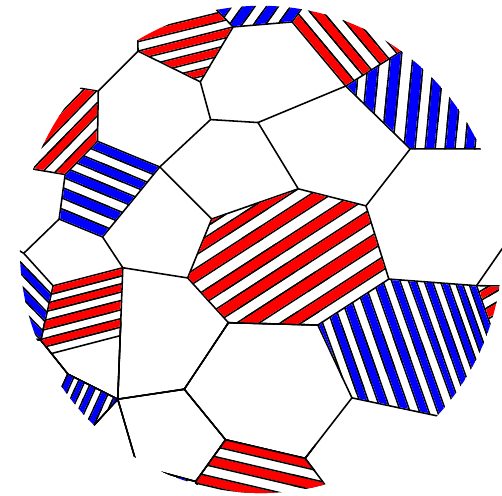}
		\caption{Schematic example of a laminate polycrystalline microstructure. Red and blue striped grains represent the stiffness obtained for different twinning configurations,
			where the habit plane of the boundary is either (011) or (101). White grains represent grains without twinning.  
		}
		\label{fig:poly_laminate_four}
	\end{figure}

	In the following, we instead focus on self-consistent estimates.
	Closed-form expressions, given as a set of implicit equations, are available  for polycrystals with cubic
	and tetragonal symmetries, see, e.g., ~\cite{berryman_bounds_2011}. 
	To take  the effect of twins into account, correlations between neighboring grains have to be taken into account, which is in general not possible for self-consistent estimates based on an isolated inclusion, known as Eshelby's inclusion, embedded in the effective medium. 
	Consider a polycrystal with a laminate substructure, as depicted in Figure~\ref{fig:poly_laminate_four}.
	This polycrystal comprises grains of two types:
	monocrystals with uniformly distributed crystallographic orientations (represented in white), 
	and grains exhibiting a laminate substructure, where the layers are in twin relation to each other (blue and red striped grains).

	The use of laminates or sequential laminates,
	for modeling polycrystalline materials is the basis of numerous works, in particular
	for constructing optimal structures that attain the Voigt and Reuss bounds~\cite{milton_theory_2002}.
	In another context, Dusthakar et al.~\cite{dusthakar_laminate-based_2018} make use of a model utilizing laminates for predicting the ferroelectric response of polycrystalline materials with  tetragonal crystal symmetry, where laminates with anisotropic layers present notable properties. 
	Goldstein et al.~\cite{goldstein_thin_2019} explored notable properties in the context of two-layered plates composed of crystals with cubic crystal symmetry. Their study revealed that for certain misorientations, the Young modulus in the normal direction surpasses that of individual plates.
	
	In the present study, we initially examine the effective elastic tensor of a rank-one laminate
	consisting of two layers denoted as $\ell_{1,2}$, where the direction, normal to the layering is denoted as $\nn$.
	The stiffness tensors $\CC_{\ell_1}$ and $\CC_{\ell_2}$ are both given with respect to the reference frame of $\ell_1$. While, this is the canonical choice for $\CC_{\ell_1}$, the tensor $\CC_{\ell_2}$ has to be transformed by applying a rotation, consistent with the twinning relationship, see Section~~\ref{subsec:twinningrel}.
	Following~\cite{mehrabadi_six-dimensional_1995,mehrabadi_eigentensors_1990} this rotation
	is obtained by a product of $6\times 6$ matrices involving the
	stiffness tensor, expressed in Voigt notation, and a $6\times 6$ \enquote{rotation} matrix.
	The stiffness tensor  of the laminate $\CC_{\ell}$
	is  given by~\cite[Eq.~7.2.7]{cherkaev_variational_2000}:
	
	\begin{eqnarray}\label{eq:lam}
		{\CC_{\ell}}^{-1}&=&
		f_1{\CC_{\ell_1}}^{-1}
		+f_2{\CC_{\ell_2}}^{-1}
		-f_1f_2\left({\CC_{\ell_2}}^{-1}-{\CC_{\ell_1}}^{-1}\right)
		TR^{-1}T^t
		\left({\CC_{\ell_2}}^{-1}-{\CC_{\ell_1}}^{-1}\right),\\
		R&=&T^t \left(f_1{\CC_{\ell_2}}^{-1}+f_2{\CC_{\ell_1}}^{-1}\right) T,\qquad
		T^t=\left( 
		\begin{tabular}{c}
			$\mathcal{Z}(\bm{t}^{(1)},\bm{t}^{(1)})$\\
			$\mathcal{Z}(\bm{t}^{(2)},\bm{t}^{(2)})$\\
			$\mathcal{Z}(\bm{t}^{(1)},\bm{t}^{(2)})$
		\end{tabular}
		\right), \nonumber \\ 
		\mathcal{Z}(\bm{v},\bm{w})&=&\left( v_1w_1; v_2w_2; v_3w_3; \frac{\sqrt{2}}{2}(v_2w_3+v_3w_2);
		\frac{\sqrt{2}}{2}(v_1w_3+v_3w_1); \frac{\sqrt{2}}{2}(v_2w_1+v_1w_2) \right),\nonumber
	\end{eqnarray}
	where $(\nn,\bm{t}^{(1)},\bm{t}^{(2)})$ is an orthogonal basis
	and $f_1$, $f_2=1-f_1$ are the volume fractions of the layers $\ell_1$ and $\ell_2$.
	The computation of the above formula involves only identifying $\CC_{\ell_1,\ell_2}$ as $6\times 6$ matrices, where $T \in \R^{6\times 3}$ and $R \in \R^{3\times 3}$.  Specify $f_1=f_2=0.5$ in Eq.~\eqref{eq:lam}, the result for the tetragonal case is given by $\CC_{\ell,\alpha}$ ($1\leq\alpha\leq\alpha_{\max}=2$) and for the cubic case, by $\CC^{C}_{\ell,\alpha}$ ($\alpha=\alpha_{\max}=1$).
	Here $\alpha_\text{max}\in\N$ denotes the number of different habit planes with respect to crystal symmetry, which are $\millerplanes{111}$ for cubic, as well as $\millerplane{011}$ and $\millerplane{101}$ for tetragonal.
	Thus, we are left with estimating the effective response of a polycrystal
	containing three (tetragonal case) or two (cubic case) types of grains, namely
	the laminates with stiffness tensors $\CC_{\ell,\alpha}$ plus the non-twined grains 
	with stiffness tensor $\CC$. The orientation of all grains is uniformly distributed on the space of crystallographic orientations and we assume all twining relationships are equiprobable, so that the 
	relative proportions of each of the two types of laminates is fixed to $0.5$ for tetragonal symmetry. 
	Subsequently a self-consistent estimate is computed by solving~\cite{willot2019thermoelastic}
	
	\begin{equation}\label{eq:scm}
		f \Biggl\langle \biggl[\Bigl({\mathbb{C}^{C,T}}^{-1}-{\mathbb{C}^*}^{-1} \Bigr)^{-1}+\mathbb{Q}\biggr]^{-1}\Biggr\rangle_{\mathcal{U}(\text{SO}_3)}
		+\frac{1-f}{\alpha_{\max}}\sum_{\alpha\leq\alpha_{\max}}    
		\Biggl\langle \biggl[\Bigl({\mathbb{C}^{C,T}_{\ell,\alpha}}^{-1}-{\mathbb{C}^*}^{-1} \Bigr)^{-1}+\mathbb{Q} \biggr]^{-1} \Biggr\rangle_{\mathcal{U}(\text{SO}_3)} = 0,
	\end{equation}
	where
	$f\in[0,1]$ is the fraction of non-twined grains and
	the averages are taken over a set of rotations, sampled from $\mathcal{U}\bigl(\text{SO}_3\bigr)$, the uniform distribution on the space of rotations $\text{SO}_3$.
	The tensor $\mathbb{Q}\in \R^{3\times 3}$ is an isotropic compliant tensor with moduli given by 
	\begin{equation*}
		q_{11}=q_{22}=q_{33}=\frac{16}{15}\cdot\frac{G}{1-\nu},\qquad
		q_{12}=q_{23}=q_{13}=\frac{2G}{15}\cdot\frac{1+5\nu}{1-\nu},
	\end{equation*}
	with the shear modulus $G$ and Poisson coefficient $\nu$ associated to $\CC^*$.\footnote{Note that isotropy of the tensor $\mathbb{Q}$ implies symmetry of its entries.}
	To solve Eq. \eqref{eq:scm}, we employ a fixed-point method. The process begins with the mean between the Voigt and Reuss bounds as the initial value, iterating until convergence is achieved.
	
	Eq.~\eqref{eq:scm} provides a self-consistent estimate for crystals with general triclinic symmetry, where the effective response is statistically isotropic. This feasibility arises because the Eshelby tensor, and consequently $\mathbb{Q}$, remains independent of the elastic tensor within the grains, relying solely on the effective tensor $\CC^*$. 
	For self-consistent estimates applicable to polycrystals with arbitrary symmetry where the effective response is isotropic, we refer to Kube and Arguelles~\cite{kube_bounds_2016}.
	
	\pgfplotsset{every tick label/.append style={font=\small},scaled y ticks=false,width=11cm,height=6.5cm}

	\subsection{Numerical results of FFT-based computations}
	For cubic crystal symmetry, the bulk modulus is the same for every configuration ($\kappa_{cubic}=\SI{110.333}{GPa}$). In contrast, the shear modulus depends on both, the grains morphology and the type of twins (Table ~\ref{table:cubic}).
	According to both,the self-consistent estimates and FFT predictions, the softest response is provided by polycrystals without twins.
	
	Numerical and analytical results for the tetragonal crystal symmetry are presented in Figure~\ref{fig:tetra_results}.
	The most significant differences between the models (tessellation model and twin configuration) can be observed in shear, with differences up to  $\SI{0.3}{GPa}$. In contrast, there is approximately no impact on the bulk modulus ($\approx \SI{1}{MPa}$). It is noteworthy that the presence of different types of twinning significantly influences the bulk modulus, which can be seen by the alignment of diamonds, circles, and triangles, each lying approximately on a horizontal line. The highest bulk modulus is attained in the absence of twins, while the lowest is achieved for inclusion twins. Note, that the type of tessellation has no significant influence on these values. Table~\ref{table:std} displays the dispersion of the values plotted on Figure~\ref{fig:tetra_results}.
	For a given twin and tessellation model, the accuracy of the FFT results for the bulk modulus is much lower than the
	differences observed between the mechanical response of the various models.
	The standard deviation for the shear modulus, however, is of the same order as the observed
	differences.

	As expected, the predictions of the self-consistent estimate~\eqref{eq:scm}
	without twins (i.e. $f=1$, end of orange line) 
	are very close to the Voronoi and Laguerre model
	in the absence of twins (red and blue diamonds).
	The presence of twins ($0<f<1$) lead to a decrease of both the effective bulk and shear moduli, consistently with FFT predictions for either neighboring or inclusion twins.

	In order to interpret these results, it is useful to 
	recall the construction of Avellaneda~\cite{avellaneda_iterated_1989},
	leading to polycrystals with the stiffest properties in terms of bulk modulus.
	A sufficient and necessary condition is that along all grain boundaries of normal $\nn$, the following relationship holds:
	
	\begin{equation}\label{eq:unia}
		\biggl\{\bm{I}:\left[\CC^{(1)}-\CC^{(2)}\right]\biggr\}\cdot{\nn} = 0,
	\end{equation}
	where $\CC^{(1)}$, $\CC^{(2)}$ are the grains' stiffness tensors, i.e.,  $\nn$
	is a null eigendirection of $\bm{I}:\bigl[\CC^{(1)}-\CC^{(2)}\bigr]$. 
	This follows from the observation that under hydrostatic loadings, a \emph{uniform} (across the two grains) strain field results in the Voigt bound.
	
	In the tetragonal case, the eigenvalues
	of $\mathbb{C}_{\ell,\alpha}^{\textnormal{T}}$
	are $[0.9;-6;9]$ and $[-1.1;-6;9]$ GPa for $\alpha=1$ and $2$, respectively.
	The lowest eigenvalue correspond to eigenvectors $(0.4,0.9,0.2)^\intercal$ and $(0.1,0.7,0.7)^\intercal$ with
	the layering directions $\bm{n}=[011]$ and $[101]$.
	The corresponding values of of the left-hand side of~\eqref{eq:unia}
	are $6$ GPa and $12$ GPa, which are far from the lowest eigenvalue. 
	\\

	\begin{table}[h]
		\begin{center}
			\begin{tabular}{c|c|c|c}
				\toprule
				& neighboring twins  & inclusion twins &  no twins \\ 
				\midrule
				Laguerre & $81.073$           & $80.964$        & $80.932$ \\ 
				Voronoi  & $81.106$           & $80.962$        & $80.944$ \\ 
				\midrule
				SC & \multicolumn{2}{c|}{$80.875$ Eq.~\eqref{eq:scm}, $f=0$}  & $80.765$
				Eq.~\eqref{eq:scm}, $f=1$ \\ 
				\bottomrule
			\end{tabular}
		\end{center}
		\caption{\label{table:cubic}
			Shear modulus $\mu$ (GPa) of both tessellation models of polycrystals with and without
			twinning, and the self-consistent estimates (SC) for cubic symmetry class. 
		}
	\end{table}

	\begin{table}[h]
		\begin{center}
			\begin{tabular}{c|cc|cc|cc}
				\toprule
				\multirow{2}{*}{} & \multicolumn{2}{c|}{neighboring twins}   & \multicolumn{2}{c|}{inclusion twins}     & \multicolumn{2}{c}{no twins}            \\ 
				& \multicolumn{1}{c}{$\kappa$}    & $\mu$       & \multicolumn{1}{c}{$\kappa$}    & $\mu$       & \multicolumn{1}{c}{$\kappa$}    & $\mu$       \\ \midrule
				Laguerre                 & \multicolumn{1}{c}{$2.28 \times 10^{-5}$} & $6.62 \times 10^{-2}$ & \multicolumn{1}{c}{$5.80 \times 10^{-6}$} & $2.36 \times 10^{-2}$ & \multicolumn{1}{c}{$4.10 \times 10^{-5}$} & $6.43 \times 10^{-2}$\\ 
				Voronoi                 & \multicolumn{1}{c}{$2.99 \times 10^{-5}$} & $6.41 \times 10^{-2}$ & \multicolumn{1}{c}{$1.28 \times 10^{-5}$} & $4.23 \times 10^{-2}$ & \multicolumn{1}{c}{$1.77 \times 10^{-5}$} & $4.99 \times 10^{-2}$ \\ 
				\bottomrule
			\end{tabular}
		\end{center}
		\caption{\label{table:std}
			Standard deviation values (in GPa) for bulk modulus $\kappa$ and shear modulus $\mu$ obtained for the three realizations of both tessellation models of polycrystals  with and without twinning for tetragonal symmetry class. 
		}
	\end{table}

	\section{Conclusion}\label{sec:conclusion}
	The stochastic 3D microstructure model for twinned polycrystals developed in this work utilizes Voronoi and Laguerre tessellations of the three-dimensional Euclidean space $\R^3$ for modeling  polycrystal grain architectures. The main advantage of these tessellation models is that they provide planar grain boundary segments which are adequate for modeling twin grains.
	Furthermore, this novel modeling approach allows for incorporating two different types of crystallographic twins, namely inclusion twins, occurring within grains, and neighboring twins, adjacent grains in a twinning relationship.
	
	Additionally, realizations of the aforementioned model served as  input for full-field Fourier-based computations  to investigate the combined effect of grain morphology and twinning on the overall elastic response. Our analysis showed that the presence of twins in $\gamma$-TiAl intermetallics is associated with a slightly softer response, especially in shear.
	This softening effect of twins has already been observed in the context of plasticity for other polycrystalline materials~\cite{clausen2008reorientation}.
	However, it is important to emphasize that this conclusion depends on the crystal symmetry and the induced twin relationship. On the other hand, it turned out that the type of tessellation (Voronoi or Laguerre) that models the morphology of the grains has less impact on the mechanical response compared to the presence of twins.
	Finally, we have derived a two-scale homogenization approach that provides predictions as a function of the fraction of neighboring twins, or inclusion twins. The prediction of these analytical estimates are consistent with full-field numerical results obtained by FFT methods. 
	
	While the  stochastic 3D model introduced in the present paper was utilized to investigate the influence of grain morphology and twinning on elastic response, it holds potential for various applications in virtual materials testing. For instance, in \cite{monteiroFernandes2024} the model was employed to analyze the influence of twins on  plastic deformation of polycrystals.
	Furthermore, beyond enabling the model to incorporate twin-related domains, 
	the development of methods for calibrating the model to grain morphology and crystallographic texture of experimentally measured image data would allow for generating virtual polycrystalline materials with twinning effect that are statistically similar to those observed by 3D imaging. This would come along with reducing the microstructural information to the values of parameters in the stochastic model and opens possibilities for the quantification of process-structure relationships for polycristalline materials with twinning, as it was done in previous studies, e.g., solar cells \cite{Westhoff2015} and battery electrodes \cite{prifling2019}.

	\section*{Acknowledgments}\label{seq:ack}
	HP thanks Safran for providing the $\gamma$-TiAl 48-2-2 material.
	The authors acknowledge  the financial support of the French Agence Nationale de la Recherche (ANR, ANR-21-FAI1-0003) and the Bundesministerium für Bildung und Forschung (BMBF, 01IS21091) within the French-German research project SMILE.

	\section*{Data availability}
	The data required to reproduce these findings will be made available to download from \\ \href{https://zenodo.org/records/10518711}{https://zenodo.org/records/10518711}.


\begin{thebibliography}{10}
	
	\bibitem{franccois2012mechanical}
	D.~Fran{\c{c}}ois, A.~Pineau, and A.~Zaoui, {\em Mechanical Behaviour of
		Materials: Volume 1: Micro-and Macroscopic Constitutive Behaviour}.
	\newblock Springer Series on Solid Mechanics and its Applications, vol. 180,
	Springer, 2012.
	
	\bibitem{groeber2008framework}
	M.~Groeber, S.~Ghosh, M.~D. Uchic, and D.~M. Dimiduk, ``A framework for
	automated analysis and simulation of {3D} polycrystalline microstructures:
	Part 1: Statistical characterization,'' {\em Acta Materialia}, vol.~56,
	1257--1273, 2008.
	
	\bibitem{klassen2012mechanical}
	M.~V. Klassen-Neklyudova, {\em Mechanical Twinning of Crystals}.
	\newblock Springer, 1964.
	
	\bibitem{redenbach2009microstructure}
	C.~Redenbach, ``Microstructure models for cellular materials,'' {\em
		Computational Materials Science}, vol.~44, 1397--1407, 2009.
	
	\bibitem{duan2014inverting}
	Q.~Duan, D.~P. Kroese, T.~Brereton, A.~Spettl, and V.~Schmidt, ``Inverting
	{Laguerre} tessellations,'' {\em The Computer Journal}, vol.~57, 1431--1440,
	2014.
	
	\bibitem{furat.2021b}
	O.~Furat, L.~Petrich, D.~P. Finegan, D.~Diercks, F.~Usseglio-Viretta, K.~Smith,
	and V.~Schmidt, ``Artificial generation of representative single li-ion
	electrode particle architectures from microscopy data,'' {\em npj
		Computational Materials}, vol.~7, 105, 2021.
	
	\bibitem{gasnier20153d}
	J.-B. Gasnier, B.~Figliuzzi, M.~Faessel, F.~Willot, D.~Jeulin, and H.~Trumel,
	``{3D} morphological modeling of a polycrystaline microstructure with
	non-convex, anisotropic grains,'' in {\em Proceedings of the 14th
		International Congress for Stereology and Image Analysis (ICSIA)}, vol.~83,
	2015.
	
	\bibitem{gasnier2015fourier}
	J.-B. Gasnier, F.~Willot, H.~Trumel, B.~Figliuzzi, D.~Jeulin, and M.~Biessy,
	``A {F}ourier-based numerical homogenization tool for an explosive
	material,'' {\em Mat{\'e}riaux \& Techniques}, vol.~103, 308, 2015.
	
	\bibitem{alpers.2015}
	A.~Alpers, A.~Brieden, P.~Gritzmann, A.~Lyckegaard, and H.~F. Poulsen,
	``Generalized balanced power diagrams for {3D} representations of
	polycrystals,'' {\em Philosophical Magazine}, vol.~95, 1016--1028, 2015.
	
	\bibitem{jung2023analytical}
	C.~Jung and C.~Redenbach, ``An analytical representation of the {2D}
	generalized balanced power diagram,'' {\em Preprint available at arXiv:
		2303.15275}, 2023.
	
	\bibitem{sedivy.2016}
	O.~{\v{S}}ediv{\`y}, T.~Brereton, D.~Westhoff, L.~Pol{\'i}vka, V.~Bene{\v{s}},
	V.~Schmidt, and A.~J{\"a}ger, ``{3D} reconstruction of grains in
	polycrystalline materials using a tessellation model with curved grain
	boundaries,'' {\em Philosophical Magazine}, vol.~96, 1926--1949, 2016.
	
	\bibitem{quey2022neper}
	R.~Quey and M.~Kasemer, ``The neper/fepx project: free/open-source polycrystal
	generation, deformation simulation, and post-processing,'' in {\em IOP
		Conference Series: Materials Science and Engineering}, vol.~1249, 012021,
	2022.
	
	\bibitem{sedivy.2018}
	O.~{\v{S}}ediv{\`y}, D.~Westhoff, J.~Kope{\v{c}}ek, C.~E. Krill~III, and
	V.~Schmidt, ``Data-driven selection of tessellation models describing
	polycrystalline microstructures,'' {\em Journal of Statistical Physics},
	vol.~172, 1223--1246, 2018.
	
	\bibitem{gasnier2018thermoelastic}
	J.-B. Gasnier, F.~Willot, H.~Trumel, D.~Jeulin, and M.~Biessy, ``Thermoelastic
	properties of microcracked polycrystals. {P}art {II}: The case of jointed
	polycrystalline {TATB},'' {\em International Journal of Solids and
		Structures}, vol.~155, 257--274, 2018.
	
	\bibitem{berryman2005bounds}
	J.~G. Berryman, ``Bounds and self-consistent estimates for elastic constants of
	random polycrystals with hexagonal, trigonal, and tetragonal symmetries,''
	{\em Journal of the Mechanics and Physics of Solids}, vol.~53, 2141--2173,
	2005.
	
	\bibitem{ambos2014a}
	A.~Ambos, H.~Trumel, F.~Willot, D.~Jeulin, and M.~Biessy, ``A fast {Fourier}
	transform micromechanical upscaling method for the study of the thermal
	expansion of a {TATB}-based pressed explosive,'' in {\em Proceedings of the
		15$^\textnormal{th}$ International Detonation Symposium}, 2014.
	
	\bibitem{willot2019thermoelastic}
	F.~Willot, H.~Trumel, and D.~Jeulin, ``The thermoelastic response of cracked
	polycrystals with hexagonal symmetry,'' {\em Philosophical Magazine},
	vol.~99, 606--630, 2019.
	
	\bibitem{willot2020elastostatic}
	F.~Willot, R.~Brenner, and H.~Trumel, ``Elastostatic field distributions in
	polycrystals and cracked media,'' {\em Philosophical Magazine}, vol.~100,
	661--687, 2020.
	
	\bibitem{clausen2008reorientation}
	B.~Clausen, C.~Tom{\'e}, D.~Brown, and S.~Agnew, ``Reorientation and stress
	relaxation due to twinning: Modeling and experimental characterization for
	{M}g,'' {\em Acta Materialia}, vol.~56, 2456--2468, 2008.
	
	\bibitem{juan2014double}
	P.-A. Juan, S.~Berbenni, M.~Barnett, C.~Tom{\'e}, and L.~Capolungo, ``A double
	inclusion homogenization scheme for polycrystals with hierarchal topologies:
	application to twinning in {M}g alloys,'' {\em International Journal of
		Plasticity}, vol.~60, 182--196, 2014.
	
	\bibitem{Kim_JOM_1991}
	Y.~Kim and D.~Dimiduk, ``Progress in the understanding of gamma titanium
	aluminides,'' {\em Journal of The Minerals, Metals \& Materials Society},
	vol.~43, 40--47, 1991.
	
	\bibitem{Dey20091052}
	S.~R. Dey, A.~Hazotte, and E.~Bouzy, ``Crystallography and phase transformation
	mechanisms in {TiAl}-based alloys -- {A} synthesis,'' {\em Intermetallics},
	vol.~17, 1052--1064, 2009.
	
	\bibitem{mccullough_phase_1989}
	C.~McCullough, J.~J. Valencia, C.~G. Levi, and R.~Mehrabian, ``Phase equilibria
	and solidification in {Ti}-{Al} alloys,'' {\em Acta Metallurgica}, vol.~37,
	1321--1336, 1989.
	
	\bibitem{kattner_thermodynamic_1992}
	U.~R. Kattner, J.~C. Lin, and Y.~A. Chang, ``Thermodynamic assessment and
	calculation of the {Ti}-{Al} system,'' {\em Metallurgical and Materials
		Transactions A}, vol.~23, 2081--2090, 1992.
	
	\bibitem{jones_phase_1993}
	S.~A. Jones and M.~J. Kaufman, ``Phase equilibria and transformations in
	intermediate titanium-aluminum alloys,'' {\em Acta Metallurgica et
		Materialia}, vol.~41, 387-- 398, 1993.
	
	\bibitem{malinov_experimental_2004}
	S.~Malinov, T.~Novoselova, and W.~Sha, ``Experimental and modelling studies of
	the thermodynamics and kinetics of phase and structural transformations in a
	gamma {TiAl}-based alloy,'' {\em Materials Science and Engineering: A},
	vol.~386, 344--353, 2004.
	
	\bibitem{TiAl_Tribeam}
	H.~Proudhon, M.~Echlin, Z.~Liu, T.~M. Pollock, and W.~Ludwig, ``3{D}
	microstructure characterization of $\gamma$-{TiAl} alloy by diffraction
	contrast tomography and {T}ri{B}eam laser sectioning,'' Working paper (in
	preparation).
	
	\bibitem{zambaldi_micromechanical_2010}
	C.~Zambaldi, {\em Micromechanical Modeling of $\gamma$-{TiAl} Based Alloys}.
	\newblock Shaker, 2010.
	
	\bibitem{Rowenhorst2015}
	D.~Rowenhorst, A.~D. Rollett, G.~S. Rohrer, M.~Groeber, M.~Jackson, P.~J.
	Konijnenberg, and M.~D. Graef, ``Consistent representations of and
	conversions between 3{D} rotations,'' {\em Modelling and Simulation in
		Materials Science and Engineering}, vol.~23, 083501, 2015.
	
	\bibitem{Echlin_MatChar_2015}
	M.~P. Echlin, M.~Straw, S.~Randolph, J.~Filevich, and T.~M. Pollock, ``The
	{T}ri{B}eam system: {F}emtosecond laser ablation in situ {SEM},'' {\em
		Materials Characterization}, vol.~100, 1--12, 2015.
	
	\bibitem{Dream3d}
	A.~Groeber and A.~Jackson, ``{DREAM.3D}: A digital representation environment
	for the analysis of microstructure in 3{D},'' {\em Integrating Materials and
		Manufacturing Innovation}, vol.~3, 5, 2014.
	
	\bibitem{Jackson_IMMI_2019}
	M.~A. Jackson, E.~Pascal, and M.~De~Graef, ``Dictionary indexing of electron
	back-scatter diffraction patterns: a hands-on tutorial,'' {\em Integrating
		Materials and Manufacturing Innovation}, vol.~8, 226--246, 2019.
	
	\bibitem{Gertsman_Scripta_1990}
	V.~Y. Gertsman, R.~M. Gayanov, A.~B. Notkin, and R.~Z. Valiev, ``Investigation
	of grain boundaries in the {TiAl} intermetallic compound,'' {\em Scripta
		Metallurgica et Materialia}, vol.~24, 1027--1032, 1990.
	
	\bibitem{chiu.2013}
	S.~Chiu, D.~Stoyan, W.~S. Kendall, and J.~Mecke, {\em Stochastic Geometry and
		Its Applications}.
	\newblock J. Wiley \& Sons, 3rd~ed., 2013.
	
	\bibitem{figliuzzi.2019}
	B.~Figliuzzi, ``Eikonal-based models of random tessellations,'' {\em Image
		Analysis \& Stereology}, vol.~38, 15--23, 2019.
	
	\bibitem{seitl.2021}
	F.~Seitl, L.~Petrich, J.~Stan{\v{e}}k, C.~E. Krill, V.~Schmidt, and
	V.~Bene{\v{s}}, ``Exploration of {Gibbs-Laguerre} tessellations for
	three-dimensional stochastic modeling,'' {\em Methodology and Computing in
		Applied Probability}, vol.~23, 669--693, 2021.
	
	\bibitem{last.2017}
	G.~Last and M.~Penrose, {\em Lectures on the Poisson Process}.
	\newblock Cambridge University Press, 2017.
	
	\bibitem{illian2008}
	J.~Illian, A.~Penttinen, H.~Stoyan, and D.~Stoyan, {\em Statistical {A}nalysis
		and {M}odelling of {S}patial {P}oint {P}atterns}.
	\newblock J. Wiley \& Sons, 2008.
	
	\bibitem{Ranganathan_ACtaCrys_1996}
	S.~Ranganathan, ``On the geometry of coincidence-site lattices,'' {\em Acta
		Crystallographica}, vol.~21, 197--199, 1966.
	
	\bibitem{Priester_2013}
	L.~Priester, {\em Grain Boundaries -- From Theory to Engineering}.
	\newblock Springer, 2013.
	
	\bibitem{Stanfill2014}
	B.~Stanfill, H.~Hofmann, and U.~Genschel, ``rotations: An {R} package for
	{SO}(3) data,'' {\em The R Journal}, vol.~6, 68--78, 2014.
	
	\bibitem{Hastie2009}
	T.~Hastie, R.~Tibshirani, and J.~Friedman, {\em The Elements of Statistical
		Learning: Data Mining, Inference, and Prediction}.
	\newblock Springer, 2009.
	
	\bibitem{moulinec_fast_1994}
	H.~Moulinec and P.~Suquet, ``A fast numerical method for computing the linear
	and nonlinear mechanical properties of composites,'' {\em Comptes Rendus de
		l'Académie des sciences. Série II}, vol.~318, 1417--1423, 1994.
	
	\bibitem{Escoda15b}
	J.~Escoda, F.~Willot, D.~Jeulin, J.~Sanahuja, and C.~Toulemonde, ``Influence of
	the multiscale distribution of particles on elastic properties of concrete,''
	{\em International Journal of Engineering Science}, vol.~98, 60--71, 2016.
	
	\bibitem{Koishi2017}
	M.~Koishi, N.~Kowatari, B.~Figliuzzi, M.~Faessel, F.~Willot, and D.~Jeulin,
	``Computational material design of filled rubbers using multi-objective
	design exploration,'' in {\em Constitutive Models for Rubber X} (A.~Lion and
	M.~Johlitz, eds.), 467--472, CRC Press, 2017.
	
	\bibitem{willot_fourier-based_2015}
	F.~Willot, ``Fourier-based schemes for computing the mechanical response of
	composites with accurate local fields,'' {\em Comptes Rendus Mécanique},
	vol.~343, 232--245, 2015.
	
	\bibitem{lantuejoul1991ergodicity}
	C.~Lantu{\'e}joul, ``Ergodicity and integral range,'' {\em Journal of
		Microscopy}, vol.~161, 387--403, 1991.
	
	\bibitem{moakher_closest_2006}
	M.~Moakher and A.~N. Norris, ``The closest elastic tensor of arbitrary symmetry
	to an elasticity tensor of lower symmetry,'' {\em Journal of Elasticity},
	vol.~85, 215--263, 2006.
	
	\bibitem{gazis_elastic_1963}
	D.~C. Gazis, I.~Tadjbakhsh, and R.~A. Toupin, ``The elastic tensor of given
	symmetry nearest to an anisotropic elastic tensor,'' {\em Acta
		Crystallographica}, vol.~16, 917, 1963.
	
	\bibitem{fedorov_theory_1968}
	F.~I. Fedorov, {\em Theory of {Elastic} {Waves} in {Crystals}}.
	\newblock Springer, 1968.
	
	\bibitem{fu_elastic_1990}
	C.~L. Fu and M.~H. Yoo, ``Elastic constants, fault energies, and dislocation
	reactions in {TiAl}: {A} first-principles total-energy investigation,'' {\em
		Philosophical Magazine Letters}, vol.~62, 159--165, 1990.
	
	\bibitem{tanaka_single-crystal_1996}
	K.~Tanaka, T.~Ichitsubo, H.~Inui, M.~Yamaguchi, and M.~Koiwa, ``Single-crystal
	elastic constants of $\gamma$-{TiAl},'' {\em Philosophical Magazine Letters},
	vol.~73, 71--78, 1996.
	
	\bibitem{berryman_bounds_2011}
	J.~G. Berryman, ``Bounds and self-consistent estimates for elastic constants of
	granular polycrystals composed of orthorhombics or crystal with higher
	symmetries,'' {\em Physical Review E}, vol.~83, 046130, 2011.
	
	\bibitem{milton_theory_2002}
	G.~W. Milton, {\em The {Theory} of {Composites}}.
	\newblock Cambridge University Press, 2002.
	
	\bibitem{dusthakar_laminate-based_2018}
	D.~K. Dusthakar, A.~Menzel, and B.~Svendsen, ``Laminate-based modelling of
	single and polycrystalline ferroelectric materials – application to
	tetragonal barium titanate,'' {\em Mechanics of Materials}, vol.~117,
	235--254, 2018.
	
	\bibitem{goldstein_thin_2019}
	R.~Goldstein, V.~Gorodtsov, D.~Lisovenko, and M.~Volkov, ``Thin homogeneous
	two-layered plates of cubic crystals with different layer orientation,'' {\em
		Physical Mesomechanics}, vol.~22, 261--268, 2019.
	
	\bibitem{mehrabadi_six-dimensional_1995}
	M.~M. Mehrabadi, S.~C. Cowin, and J.~Jaric, ``Six-dimensional orthogonal
	tensorrepresentation of the rotation about an axis in three dimensions,''
	{\em International Journal of Solids and Structures}, vol.~32, 439--449,
	1995.
	
	\bibitem{mehrabadi_eigentensors_1990}
	M.~Mehrabadi and S.~Cowin, ``Eigentensors of linear anisotropic elastic
	materials,'' {\em The Quarterly Journal of Mechanics and Applied
		Mathematics}, vol.~43, 15--41, 1990.
	
	\bibitem{cherkaev_variational_2000}
	A.~Cherkaev, {\em Variational {Methods} for {Structural} {Optimization}}.
	\newblock Springer, 2000.
	
	\bibitem{kube_bounds_2016}
	C.~M. Kube and A.~P. Arguelles, ``Bounds and self-consistent estimates of the
	elastic constants of polycrystals,'' {\em Computers and Geosciences},
	vol.~95, 118--122, 2016.
	
	\bibitem{avellaneda_iterated_1989}
	M.~Avellaneda, ``Iterated homogenization and the effective properties of
	polycrystals,'' in {\em Control of Boundaries and Stabilization} (J.~Simon,
	ed.), Lecture Notes in Control and Information Sciences, vol. 125, 66--74,
	Springer, 1989.
	
	\bibitem{monteiroFernandes2024}
	L.~Monteiro~Fernandes, P.~Rieder, M.~Neumann, A.~Mulard, H.~Proudhon,
	V.~Schmidt, and F.~Willot, ``Effect of crystallographic twins on the
	elastoplastic response of polycrystals,'' {\em Preprint available at arXiv:
		2402.09996}, 2024.
	
	\bibitem{Westhoff2015}
	D.~Westhoff, J.~J. van Franeker, T.~Brereton, D.~P. Kroese, R.~A.~J. Janssen,
	and V.~Schmidt, ``Stochastic modeling and predictive simulations for the
	microstructure of organic semiconductor films processed with different spin
	coating velocities,'' {\em Modelling and Simulation in Materials Science and
		Engineering}, vol.~23, 045003, 2015.
	
	\bibitem{prifling2019}
	B.~Prifling, D.~Westhoff, D.~Schmidt, H.~Markötter, I.~Manke, V.~Knoblauch,
	and V.~Schmidt, ``Parametric microstructure modeling of compressed cathode
	materials for {L}i-ion batteries,'' {\em Computational Materials Science},
	vol.~169, 109083, 2019.
	
\end{thebibliography}
\end{document}